\pdfoutput=1
\documentclass[12pt, onecolumn, draftcls]{IEEEtran}
\usepackage{amsmath,amssymb}
\usepackage[dvips]{graphicx}
\usepackage{amsfonts}
\usepackage[mathscr]{eucal}
\usepackage{latexsym}
\usepackage{amsthm}
\usepackage{exscale}
\usepackage[mathscr]{eucal}
\usepackage{bm}
\usepackage[dvipsnames]{color}
\usepackage{cases}
\usepackage{color}
\usepackage{epsfig}
\usepackage{algorithm}
\usepackage{algorithmic}
\usepackage[verbose,nospace,sort]{cite}
\usepackage{tabularx}
\usepackage{multirow}
\usepackage{multicol}
\usepackage{balance}
\usepackage{caption}
\usepackage{subcaption}
\usepackage{setspace}
\usepackage{url}

\scrollmode

\theoremstyle{definition}

\newcommand{\thr}{\mathrm{th}}

\newcommand{\out}{\mathrm{out}}
\newcommand{\SIR}{\mathrm{SIR}}

\usepackage{amsmath}

\begin{document}
\title{Simultaneous Navigation and Radio Mapping\\
 for Cellular-Connected UAV \\
 with Deep Reinforcement Learning}
\author{Yong Zeng, Xiaoli Xu, Shi Jin, and Rui Zhang  \vspace{-4ex}\\
\thanks{Y. Zeng, X. Xu, and S. Jin are with the National Mobile Communications Research Laboratory, Southeast University, Nanjing 210096, China. Y. Zeng is also with Purple Mountain Laboratories, Nanjing 211111, China (e-mail: \{yong\_zeng, xiaolixu, jinshi\}@seu.edu.cn).}
\thanks{R. Zhang is with the Department of Electrical and Computer Engineering, National University of Singapore (e-mail: elezhang@nus.edu.sg).}
\thanks{Part of this work has been presented in IEEE Global Communications Conference, 9-13 December 2019, Waikoloa, HI, USA \cite{1101}.}
}

\maketitle

\begin{abstract}
Cellular-connected unmanned aerial vehicle (UAV) is a promising technology to unlock the full potential of UAVs in the future by reusing the cellular base stations (BSs) to enable their air-ground communications. However, how to achieve ubiquitous three-dimensional (3D) communication coverage for the UAVs in the sky is a new challenge. In this paper, we tackle this challenge by a new  \emph{coverage-aware navigation} approach, which exploits the UAV's controllable mobility to design its navigation/trajectory to avoid the cellular BSs' coverage holes while accomplishing their missions. To this end, we formulate an UAV trajectory optimization problem to minimize the weighted sum of its mission completion time and expected communication outage duration, which, however, cannot be solved by the standard optimization techniques mainly due to the lack of an accurate and tractable end-to-end communication model in practice. To overcome this difficulty, we propose a new solution approach  based on the technique of \emph{deep reinforcement learning} (DRL). Specifically, by leveraging the state-of-the-art \emph{dueling double deep Q network (dueling DDQN) with multi-step learning}, we first propose a UAV navigation algorithm based on direct RL, where the signal measurement at the UAV is used to directly train the \emph{action-value function} of the navigation policy. To further improve the performance, we propose a new framework called \emph{simultaneous navigation and radio mapping (SNARM)}, where the UAV's signal measurement is used not only for training the DQN directly, but also to create a radio map that is able to predict the outage probabilities at all locations in the area of interest. This thus enables the generation of simulated UAV trajectories and predicting their expected  returns, which are then  used to further train the DQN  via \emph{Dyna} technique, thus greatly improving the learning efficiency. Simulation results demonstrate the effectiveness of the proposed algorithms for coverage-aware UAV navigation, as well as the significantly improved performance of SNARM over direct RL.
\end{abstract}
%

\section{Introduction}
Conventionally, cellular networks are designed  to  mainly serve terrestrial user equipments (UEs) with fixed infrastructure. 
With the continuous expansion of human activities towards the sky and the fast growing use of unmanned aerial vehicles (UAVs) for various applications, there have been increasing interests in integrating UAVs into cellular networks \cite{1095}. On one hand, dedicated UAVs could be dispatched as aerial base stations (BSs) or relays to assist wireless communications between devices without direct connectivity, or  as flying access points (APs) for data collection and information dissemination, leading to a new three-dimensional (3D) network with {\it UAV-assisted communications} \cite{649}. On the other hand, for those UAVs with their own missions,  cellular network could be utilized to support their command and control (C\&C) communication as well as payload data transmissions, corresponding to the other paradigm of {\it cellular-connected UAV} \cite{952}. 
In particular, by reusing the existing densely deployed cellular BSs worldwide, together with the advanced cellular technologies, cellular-connected UAV has the great potential to achieve truly remote UAV operation with unlimited range, not to mention other advantages such as the ease of legitimate UAV monitoring and management, high-capacity payload transmission, and cellular-enhanced positioning \cite{1095,952}.


However, to practically realize the above vision of cellular-connected UAVs,  there are still several critical challenges to be addressed. In particular, as existing cellular networks are mainly planned for ground coverage with BS antennas typically downtilted towards the ground \cite{1024}, unlike that on the ground,  ubiquitous cellular coverage in the sky cannot be guaranteed in general \cite{941}. In fact, even for the forthcoming 5G or future 6G cellular networks that are planned to  embrace the new type of aerial UEs, it is still unclear whether ubiquitous sky coverage is economically viable or not, even for some moderate range of altitude, considering various practical factors  such as infrastructural and operational costs, as well as the anticipated aerial UE density in the near future. Such coverage issue is exacerbated by the more severe aerial-ground interference as compared to the terrestrial counterpart \cite{941,1012,952}, due to the high likelihood of having strong line of sight (LoS) channels for the UAVs with their non-associated co-channel BSs.

 Fortunately, different from the conventional ground UEs whose mobility is typically random and uncontrollable, which renders ubiquitous ground coverage essential, the mobility of UAVs is more predictable and most of the time controllable, either autonomously by computer program or by remote pilots. This  thus offers a new degree of freedom to circumvent the aforementioned coverage issue, via {\it coverage-aware navigation or trajectory design}: an approach that requires no or little modifications for cellular networks to serve aerial UEs. There are some preliminary efforts along this direction. In \cite{1080}, by applying graph theory and convex optimization for cellular-connected UAV, the UAV trajectory is optimized to minimize the UAV travelling time while ensuring that it is always connected with at least one BS. A similar problem is studied in \cite{1198,1008}, by allowing  certain tolerance for disconnection.

 However, the conventional optimization-based UAV trajectory design like those mentioned above have practical limitations. First, formulating the corresponding optimization problems requires accurate and analytically tractable end-to-end communication models, including the BS antenna model, channel model, interference model and even local environmental model. Perhaps for such reasons,  prior works such as \cite{1080,1198,1008} have mostly assumed  some simplified models like isotropic radiation for antennas and/or free-space path loss channel model for aerial-ground links. Although there have been previous works considering more sophisticated channel models, like the probabilistic LoS channel model \cite{642} and angle-dependent channel parameters \cite{954,1086}, these are statistical models that can only predict the performance in the average sense without providing any performance guarantee for the local environment where the UAVs are actually deployed.  Another limitation of off-line optimization-based trajectory design is the requirement of perfect and usually global knowledge of the channel model parameters, which is difficult to acquire in practice. Last but not least, even with the accurate channel model and the perfect information of all relevant parameters, most of these off-line optimization problems are highly non-convex and thus difficult to be efficiently solved.

\begin{figure}
\centering
\includegraphics[width=0.4\textwidth]{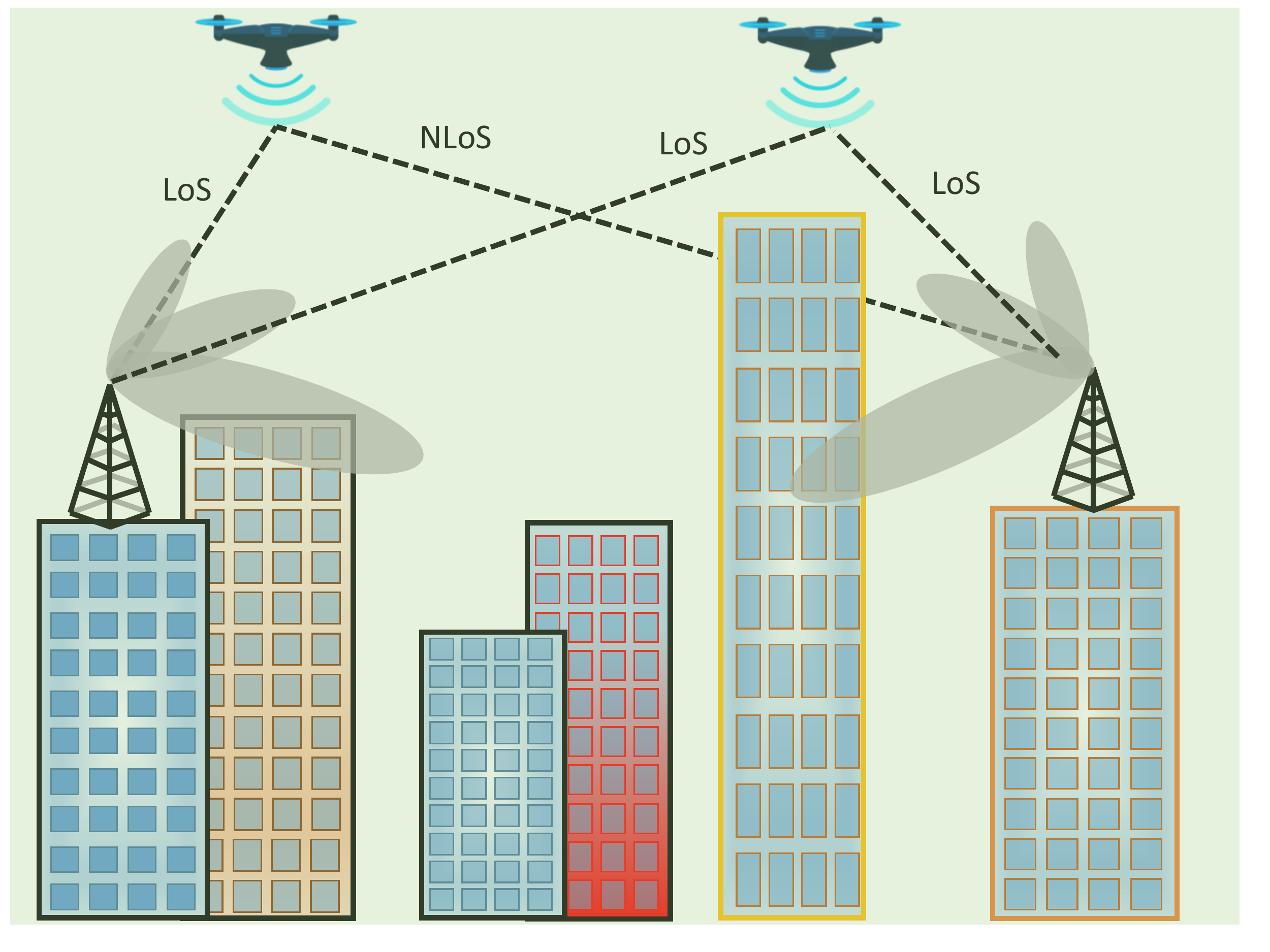}
\caption{Coverage-aware navigation for cellular-connected UAVs in urban environment.\vspace{-3ex}}
\label{F:SystemModel}
\end{figure}

To overcome the above limitations, we propose in this paper a new approach for coverage-aware UAV navigation based on reinforcement learning (RL) \cite{1084}, which is one type of machine learning techniques for solving sequential decision problems. As shown in Fig.~\ref{F:SystemModel}, we consider a typical cellular-connected UAV system in urban environment, where multiple UAVs need to operate in a given airspace with their communications supported by cellular BSs. The UAV missions are to fly from their respective initial locations to a common final location with minimum time,\footnote{The results of this paper can be extended to the general case of different final locations for the UAVs.} while maintaining a satisfactory communication connectivity with the cellular network with best effort. Our main contributions are summarized as follows:

 \begin{itemize}
 \item First, we formulate the UAV trajectory optimization problem to minimize the weighted sum of its mission completion time and the expected communication outage duration, which, however, is difficult to be solved via the standard optimization techniques mainly due to the lack of an accurate and tractable communication model. Fortunately, we show that the formulated problem can be transformed to an equivalent Markov decision process (MDP), for which a handful of RL techniques, such as the classic Q learning \cite{1084}, can be applied to learn the UAV action policy, or the flying direction in our problem.
 \item As the resulting MDP involves continuous state space that essentially has infinite number of state-action pairs and thus renders the simple table-based Q learning inapplicable, we apply the celebrated deep neural network (DNN) to approximate the {\it action-value function} (or Q function), a technique commonly known as deep reinforcement learning (DRL) or more specifically, deep Q learning \cite{1186}. To train the DNN, we use the state-of-the-art dueling network architecture with double DQN (dueling DDQN) \cite{1187} and multi-step learning. Different from the conventional off-line optimization-based designs, the proposed DRL-based trajectory design does not require any prior knowledge about the channel model or propagation environment, but only utilizes the raw signal measurement at each UAV as the input to improve its radio environmental awareness, as evident from the continuously  improved estimation accuracy of the UAVs' action-value functions.
 \item As direct RL based on real experience or measured data only is known to be data inefficient, i.e., a large amount of data or agent-environment interaction is needed, we further propose a new design framework termed {\it simultaneous navigation and radio mapping (SNARM)} to improve the learning efficiency. Specifically, with SNARM, the network (say, a macrocell BS in the area of interest) maintains a database, also known as {\it radio map} \cite{1062,1067}, which provides the outage probability distribution over the space of interest with respect to the relatively stable (large-scale) channel parameters. As such, the obtained signal measurements as UAVs fly along their trajectories are used for dual purposes: to train the DQN to improve the estimation of the optimal action-value functions, as well as to improve the accuracy of the estimated radio map, which is used for the prediction of the outage probability for other locations in the considered area even if they have not been visited by any UAV yet. The exploitation of the radio map, whose accuracy is continuously improved as more signal measurements are accumulated, makes it possible to generate simulated UAV trajectories  and predict their corresponding returns, without having to actually fly the UAVs along them. This thus leads to our proposed SNARM framework, by applying the classic {\it Dyna} \cite{1084} dueling DDQN with multi-step learning to train the action-value functions with a combined use of real trajectories and simulated trajectories, thus greatly reducing the actual  UAV flights or measurement data required.
 \item Numerical results are provided, with the complete source codes available online\footnote{\url{https://github.com/xuxiaoli-seu/SNARM-UAV-Learning}}, to verify the performance of the proposed algorithms. It is observed that the SNARM algorithm is able to learn the radio map very efficiently, and both direct RL and SNARM algorithms are able to effectively navigate UAVs to avoid the weak coverage regions of the cellular network. Furthermore, thanks to the simultaneous learning of the radio map, SNARM is able to significantly reduce the number of required UAV flights (or learning episodes) than direct RL while achieving comparable performance.
 \end{itemize}

Note that applying RL techniques for UAV communications has received significant attention recently \cite{1057,1059,cui2018multi,1188,1202} and DRL has also been  used in wireless networks \cite{1189} for solving various problems like multiple access control \cite{1190}, power allocation \cite{1200}, modulation and coding \cite{1201}, mobile edge computing \cite{1191}, UAV BS placement \cite{1192}, and UAV interference management \cite{1060}, among others. 
On the other hand, local environmental-aware UAV path/trajectory design has been studied from different perspectives, such as that based on the known radio map \cite{shuowenzhang2019} or 3D city map \cite{1066,1194}, or the joint online and off-line optimization for UAV-enabled data harvesting \cite{1193}. Compared to such existing path/trajectory designs, the proposed coverage-aware navigation algorithms based on direct RL or SNARM do not require any prior knowledge or assumption about the environment, but only utilize the UAV's measured data as the input for action and/or radio map learning, thus  expected to be more robust to any imperfect knowledge or change of the local environment. Although large data measurements are generally required for the proposed algorithms, they are practically available according to the existing cellular communication protocols, such as the reference signal received power (RSRP) and reference signal received quality (RSRQ) measurements. Besides, as different UAVs in the same region share the same local radio propagation environment, regardless of operating concurrently or at different time, their measured data as they perform their respective missions can actually be collectively utilized for our proposed designs, to continuously improve the quality of the learning for the future missions, which thus greatly alleviates the burden for data acquisition in practice.

The rest of the paper is organized as follows. Section~\ref{sec:SystemModel} introduces the system model and problem formulation. In Section~\ref{sec:DRLOverview}, we give a very brief background about RL and DRL, mainly to introduce the key concepts and main notations. Section~\ref{sec:ProposedAlgo} presents the proposed algorithms, including the direct RL based on dueling DDQN with multi-step learning for UAV navigation, and SNARM to enable path learning with both real and simulated flight experience. Section~\ref{sec:Numerical} presents the numerical results, and finally we conclude the paper in Section~\ref{sec:Conclusion}.

\section{System Model and Problem Formulation}\label{sec:SystemModel}
\begin{figure}
\centering
\includegraphics[width=0.4\textwidth]{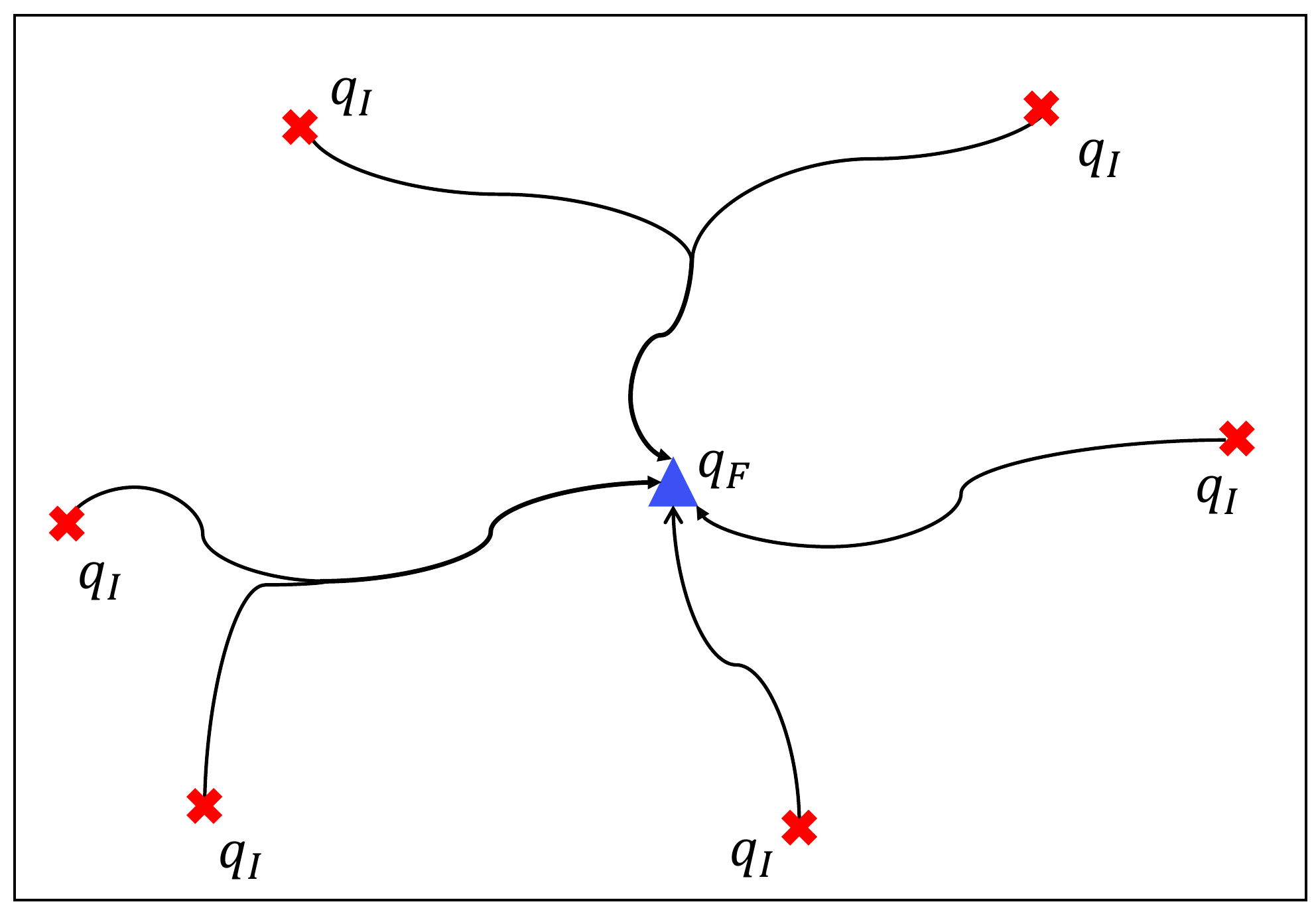}
\caption{UAVs with various initial locations fly to a common destination.\vspace{-3ex}}
\label{F:trajCommonDest}
\end{figure}

As shown in Fig.~\ref{F:SystemModel}, we consider a cellular-connected UAV system with arbitrary number of UAV users. Without loss of generality, we assume that the airspace of interest is a cubic volume, which is specified by $[x_L, x_U] \times [y_L, y_U] \times [z_L, z_U]$, with the subscripts $L$ and $U$ respectively denoting the lower and upper bounds of the 3D coordinate of the airspace. UAVs with various missions need to operate in this airspace with communications supported by cellular BSs. While different UAVs usually have different missions and may  operate at different time, since they share the common airspace, their flight experience including their signal measurements along their respective flying trajectories, can be exploited collectively to improve their radio environmental awareness. To make the problem concrete, we consider the scenario that the UAVs need to fly from their respective initial locations $\mathbf q_I\in \mathbb{R}^{3\times 1}$, which are generally different for different UAVs, to a common final location $\mathbf q_F\in \mathbb{R}^{3\times 1}$, as shown in Fig.~\ref{F:trajCommonDest}. This setup corresponds to various UAV applications such as UAV-based parcel collection, with $\mathbf q_I$ corresponding to the customer's home location and $\mathbf q_F$ being the local processing station of courier company, or UAV-based aerial inspection, with $\mathbf q_I$ being the target point for inspection and $\mathbf q_F$ being the charging station in the local area, etc. We assume that collision/obstacle avoidance is guaranteed by e.g., ensuring that UAVs are sufficiently separated in operation time or space, and their flight altitudes are higher than the maximum building height. Without loss of generality, we can focus on one typical UAV, since other UAVs may apply similar strategies to update the common location-dependent database, such as the action-value function of the RL algorithms and the radio map of the environment (to be defined later). The UAV needs to be navigated from its initial location $\mathbf q_I$ to the final location $\mathbf q_F$ with the minimum mission completion time, while maintaining a satisfactory communication connectivity with the cellular network. Let $T$ denote the mission completion time and $\mathbf q(t)\in \mathbb{R}^{3\times 1}$, $t\in [0, T]$, represent the UAV trajectory. Then we should have
\begin{align}
& \mathbf q(0) =\mathbf q_I,\ \mathbf q(T)=\mathbf q_F, \\
& \mathbf q_L\preceq\mathbf q(t)\preceq \mathbf q_U,  \forall t\in (0, T), \label{eq:trajconstr3}
\end{align}
where $\mathbf q_L\triangleq [x_L;y_L;z_L]$ and $\mathbf q_U\triangleq[x_U;y_U;z_U]$, with $\preceq$ denoting the element-wise inequality. 

Let $M$ denote the total number of cells in the area of interest, and $h_m(t)$, $1\leq m \leq M$, represent the baseband equivalent channel from cell $m$ to the UAV at time $t$, which generally depends on the BS antenna gains, the large-scale path loss and shadowing that are critically dependent on the local environment, as well as  the small-scale fading. The received signal power at time $t$ by the UAV from cell $m$ can be written as
\begin{align}
p_m(t)&=\bar P_m |h_m(t)|^2\\
&=\bar P_m \beta_m (\mathbf q(t)) G_m(\mathbf q(t))\tilde h_m(t), \ m=1,\cdots, M, \label{eq:pmt}
\end{align}
where $\bar P_m$ denotes the transmit power of cell $m$ that is assumed to be a constant, $\beta_m(\cdot)$ and $G_m(\cdot)$ denote the large-scale channel gain and the BS antenna gain, respectively, which generally depend on the UAV location $\mathbf q(t)$, and $\tilde h_m(t)$ is a random variable accounting for the small-scale fading. As the proposed designs in this paper work for arbitrary channels, a detailed discussion of one particular BS-UAV channel model adopted for our simulations is deferred to Section~\ref{sec:Numerical}.

Denote by $b(t)\in \{1,\cdots, M\}$ the  cell associated to the UAV at time $t$. We say that the UAV is in outage at time $t$ if its received signal-to-interference ratio (SIR) is below a certain threshold $\gamma_{\thr}$, i.e., when  $\SIR(t)<\gamma_{\thr}$,
where
\begin{align}
\SIR(t)\triangleq \frac{p_{b(t)}(t)}{\sum_{m\neq b(t)} p_m(t)}.\label{eq:gammat}
\end{align}
Note that for simplicity, we have ignored the noise effect, since BS-UAV communication is known to suffer from more severe interference than the terrestrial counterparts, and thus its communication performance is usually interference-limited. Besides, we consider the worst-case scenario with universal frequency reuse so that all these non-associated BSs would contribute to the sum-interference term in \eqref{eq:gammat}. The extension of the proposed techniques for dynamic interference will be an interesting work for further investigation.

Due to the randomness of small-scale fading, for any given UAV location $\mathbf q(t)$ and cell association  $b(t)$ at time $t$, $\SIR(t)$ in \eqref{eq:gammat} is a random variable, and the resulting outage probability is a function of $\mathbf q(t)$ and $b(t)$, i.e.,
\begin{align}\label{eq:Pout}
P_\out(\mathbf q(t),b(t))\triangleq \Pr\left(\SIR(t)<\gamma_{\thr}\right),
\end{align}
where $\Pr\left(\cdot\right)$ is the probability of the event taken with respect to the randomness of small-scale fading $\{\tilde h_m(t)\}_{m=1}^M$. 
Then the total expected outage duration across the UAV's mission duration $T$ is
\begin{align}
\bar T_\out(\{\mathbf q(t),b(t)\})=\int_0^T P_\out(\mathbf q(t),b(t)) dt. \label{eq:barTout}
\end{align}


Intuitively, with longer mission completion time $T$, the UAV is more flexible to adapt its trajectory to avoid the weak coverage regions of the cellular network in the sky and thus reduce the expected outage duration $\bar T_{\out}$. Therefore, there in general exists a tradeoff between minimizing the mission completion time $T$ and expected outage duration $\bar T_{\out}$, which can be balanced by designing  $\{\mathbf q(t)\}$ and $\{b(t)\}$ to minimize the weighted sum of these two metrics with certain weight $\mu \geq 0$, as formulated in the following optimization problem.
\begin{align}
\mathrm{(P0):} \underset{T, \{\mathbf q(t),b(t)\}}{\min}   \ & T + \mu \bar T_{\out}(\{\mathbf q(t),b(t)\}) \notag \\
\mathrm{s.t.} \  &\|\dot{\mathbf q}(t)\| \leq V_{\max}, \ \forall t\in [0, T], \\
& \mathbf q(0)=\mathbf q_I, \ \mathbf q(T)=\mathbf q_F, \label{eq:qFConstr}\\
& \mathbf q_L \preceq \mathbf q(t) \preceq \mathbf q_U, \ \forall t\in [0,T], \label{eq:RectConstr}\\
& b(t) \in \{1,\cdots, M\},\label{eq:binary}
\end{align}
where $V_{\max}$ denotes the maximum UAV speed. It can be shown that at the optimal solution to $\mathrm{(P0)}$, the UAV should fly with the maximum speed $V_{\max}$, i.e., $\dot{\mathbf q}(t)=V_{\max} \vec{\mathbf v}(t)$, where $\vec{\mathbf v}(t)$ denotes the UAV flying direction with $\|\vec{\mathbf v}(t)\|=1$. Thus, $\mathrm{(P0)}$ can be equivalently written as
\begin{align}
\mathrm{(P1):} \underset{T, \{\mathbf q(t), \vec{\mathbf v}(t),b(t)\}}{\min}   \ & T + \mu \bar T_{\out}(\{\mathbf q(t),b(t)\}) \notag \\
\mathrm{s.t.} \  &\dot{\mathbf q}(t) = V_{\max}\vec{\mathbf v}(t), \ \forall t\in [0, T], \label{eq:DifferentialEq}\\
& \|\vec{\mathbf v}(t)\|=1, \ \forall t\in [0, T], \label{eq:NormConstr}\\
& \eqref{eq:qFConstr}-\eqref{eq:binary}. \notag
\end{align}

In practice, finding the optimal UAV trajectory and cell association  by directly solving the optimization problem $\mathrm{(P1)}$ faces several challenges. First, to complete the problem formulation of $\mathrm{(P1)}$ to make it ready for optimization, we still need to derive the closed-form expression for the expected outage duration $\bar T_{\out}(\{\mathbf q(t),b(t)\})$ for any given UAV location $\mathbf q(t)$ and association $b(t)$, which is challenging, if not impossible. On one hand, this requires accurate and tractable modelling of the end-to-end channel $h_m(t)$ for cellular-UAV communication links, where practical 3D BS antenna pattern \cite{1106} and the frequent variations of LoS/NLoS connections, which depends on the actual building/terrain distributions in the area of interest, need to be considered. On the other hand, even with such modelling and the perfect knowledge of the terrain/building information for the area of interest, deriving the closed-form expression for $\bar T_{\out}(\{\mathbf q(t),b(t)\})$ is mathematically difficult since the probability density function (pdf) of the SIR depends on the UAV trajectory $\mathbf q(t)$  in sophisticated manners, as can be inferred from \eqref{eq:pmt} and \eqref{eq:gammat}. Furthermore, even after completing the problem formulation of $\mathrm{(P1)}$ with properly defined cost function, the resulting optimization problem will be highly non-convex and difficult  to be efficiently solved.

To overcome the above issues, in the following, we propose a novel approach for coverage-aware UAV navigation by leveraging the powerful mathematical framework  of RL and DNN, which only requires the UAV signal measurements along its flight as the input. By leveraging the state-of-the-art dueling DDQN with multi-step learning \cite{1197}, we first propose the direct RL-based algorithm for coverage-aware UAV navigation, and then introduce the more advanced SNARM algorithm to improve the learning efficiency and UAV performance.
\vspace{-0.1cm}
\section{Overview of Deep Reinforcement Learning}\label{sec:DRLOverview}
This section aims to give a brief overview on RL and DRL, by introducing the key notations to be used in the sequel of this paper. The readers are referred to the classic textbook \cite{1084} for a more comprehensive description.
\vspace{-0.1cm}
\subsection{Basics of Reinforcement Learning}\label{sec:RLOverview}
RL is a useful machine learning method to solve the MDP \cite{1084}, which consists of an {\it agent} interacting with the {\it environment} iteratively. For the fully observable MDP, at each discrete time step $n$, the agent observes a state $S_n$, takes an action $A_n$, and then receives an immediate reward $R_n$ and  transits to the next state $S_{n+1}$. Mathematically, an MDP can be specified by 4-tuple $<\mathcal S, \mathcal A, \mathcal P, \mathcal R>$, where
$\mathcal S$ is the state space, $\mathcal A$ is the action space, $\mathcal P$ is the state transition probability, with $P(s^\prime | s, a)$ specifying the probability of transiting to the next state  $s^\prime\in \mathcal S$ given the current state $s\in \mathcal S$ after applying the action $a\in \mathcal A$, and $\mathcal R$ is  the immediate reward received by the agent, usually denoted by $R_n$ for reward received at time step $n$ or $R(s,a)$ to show its general dependency on $s$ and $a$.


The agent's actions are governed by its policy $\pi: \mathcal S \times \mathcal A \rightarrow [0,1]$, where $\pi(a|s)$ gives the probability of taking action $a\in \mathcal A$ when in state $s\in \mathcal S$. 
 The  goal of the agent is to improve its policy $\pi$ based on its experience, so as to maximize its long-term expected {\it return} $\mathbb{E}[G_n]$, where the return $G_n\triangleq \sum_{k=0}^{\infty} \gamma^k R_{n+k}$ is the accumulated discounted  reward from time step $n$ onwards with a discount factor $0\leq \gamma \leq 1$. 

A key metric of RL is the {\it action-value function}, denoted as $Q_{\pi}(s,a)$, which is the expected return starting from state $s$, taking the action $a$, and following policy $\pi$ thereafter, i.e.,
$Q_{\pi}(s,a)=\mathbb E_\pi [G_n| S_n=s, A_n=a]$.
The optimal action-value function is defined as $Q_*(s,a)=\underset{\pi}{\max}\  Q_\pi(s,a)$, $\forall s\in \mathcal S$ and $a\in \mathcal A$.
If the optimal value function $Q_*(s,a)$  is known, the optimal policy $\pi_*(a|s)$ can be directly obtained as   
\begin{align}
\pi_*(a|s)=\begin{cases} 1, &\text{ if } a=\underset{a\in \mathcal A}{\mathrm{argmax}} \ Q_*(s,a), \\
0, & \text{ otherwise. }
\end{cases}
\end{align}
Thus, an essential task of  RL  is to obtain the optimal value functions, which satisfy the celebrated Bellman optimality equation \cite{1084}
\begin{align}
Q_*(s,a)=R(s,a)+ \gamma\ \sum_{s^\prime \in \mathcal S} P(s'|s,a)\underset{a'\in \mathcal A}{\max}\ Q_*(s',a'),  \forall s\in \mathcal S, a \in \mathcal A.
\end{align}
Bellman optimality equation is non-linear, and admits no closed-form solution in general. However, many iterative solutions have been proposed. Specifically, when the agent has the full knowledge of the MDP, including the reward $\mathcal R$ and the state-transition probabilities $\mathcal P$, the optimal value function can be obtained recursively based on dynamic programming, such as {\it value iteration} \cite{1084}.
On the other hand, when the agent has no or incomplete prior knowledge about the MDP, it may apply the useful idea of {\it temporal difference} (TD) learning to improve its estimation of the value function by directly interacting with the environment. Specifically, TD learning is a class of {\it model-free} RL methods that learn the value functions based on the direct samples of the {\it state-action-reward-nextState} sequence, with the estimation of the value functions updated by the concept of {\it bootstrapping}. One classic TD learning method is {\it Q-learning}, which applies the following update to the action-value function with the observed sample reward-and-state transitions $(S_n, A_n, R_n, S_{n+1})$:
\begin{equation}
\begin{aligned}
\hspace{-4ex} Q(S_n, A_n)\leftarrow Q(S_n, A_n) + \alpha \left[ R_n + \gamma\ \underset{a\in \mathcal A}{\max}\  Q (S_{n+1},a)-Q(S_n,A_n)\right],\notag
\end{aligned}
\end{equation}
where $\alpha$ is the learning rate. It has been shown that Q-learning is able to converge to the optimal action-value function $Q_*$ if each state-action pair is visited by sufficient times and appropriate learning rate $\alpha$ is chosen \cite{1084}.

\vspace{-0.1cm}
\subsection{Deep Reinforcement Learning}\label{sec:DRL}
The RL learning method discussed in Section~\ref{sec:RLOverview} is known as {\it table-based}, which requires storing one value for each state-action pair, and the value is updated only when that state-action pair is actually experienced. This  becomes impractical for continuous state/action or when the number of discretized states/actions is very large. In order to practically apply RL algorithms to large/continuous state-action space, one may resort to the useful technique of {\it function approximation} \cite{1084}, where the action-value function is approximated by certain parametric function, e.g., $Q(s,a)\approx \hat Q(s, a;\boldsymbol \theta)$, $\forall s\in \mathcal S, a\in \mathcal A$, with a parameter vector $\boldsymbol \theta$. Function approximation brings two advantages over table-based RL. Firstly, instead of storing and updating the value functions for all state-action pairs, one only needs to learn the parameter $\boldsymbol \theta$, which typically has much lower dimension than the number of state-action pairs. Secondly, function approximation enables {\it generalization}, i.e., the ability to predict the values even for those state-action pairs that have never been experienced, since different state-action pairs are coupled with each other via the function $\hat Q(s, a;\boldsymbol \theta)$ and parameter $\boldsymbol \theta$. 

One powerful non-linear function approximation technique is via using artificial neural networks (ANNs), which are networks consisting of interconnected units (or neurons), with each unit computing a weighted sum of their input signals and then applying a nonlinear function (called the {\it activation function}) to produce its output. It has been theoretically shown that an ANN with even a single hidden layer containing a sufficiently large number of neurons is a universal function approximation \cite{1084}, i.e., it can approximate any continuous function on a compact region to any degree of accuracy. Nonetheless, researchers have found that ANNs with deep architectures consisting of many hidden layers are of more practical usage for complex function approximations. A combination of RL with function approximation based on deep ANNs leads to the powerful framework of DRL \cite{1186}.

With deep ANN for the action-value function approximation $Q(s,a)\approx \hat Q(s, a;\boldsymbol \theta)$, the parameter vector $\boldsymbol \theta$ actually corresponds to the weight coefficients and bias of all links in the ANNs. In principle, $\boldsymbol \theta$ can be updated based on the classic backpropagation algorithm \cite{1199}. However, its application for DRL faces new challenges, since standard ANNs are usually trained in the {\it supervised learning} manner, i.e., the labels (in this case the true action values) of the input training data are known, which is obviously not the case for DRL. This issue can be addressed by the idea of {\it bootstrapping}, i.e., for each given {\it state-action-reward-nextState} transition $(S_n, A_n, R_n, S_{n+1})$, the parameter of the network $\boldsymbol \theta$ is updated to minimize the loss given by 
\begin{align}
\left(R_n +\gamma \underset{a\in \mathcal A}{\max}\ \hat Q (S_{n+1},a; \boldsymbol \theta)-\hat Q(S_n, A_n;\boldsymbol \theta) \right)^2. \label{eq:target1}
\end{align}
However, as the target in \eqref{eq:target1} also depends on the parameter $\boldsymbol \theta$  to be actually updated, directly applying the standard deep training algorithms with \eqref{eq:target1} can lead to oscillations or even divergence. This problem was addressed in the celebrated work \cite{1186}, which proposed the simple but effective technique of ``target network'' to bring the Q-learning update closer to the standard supervised-learning setup. A target network, denoted as $\hat Q(s,a;\boldsymbol \theta^-)$ with parameter $\boldsymbol \theta^-$, is a copy of the Q network $\hat Q(s,a;\boldsymbol \theta)$, but with its parameter $\boldsymbol \theta^-$ updated much less frequently. For example, after every $B$ number of updates to the weights $\boldsymbol \theta$ of the Q network, we may set $\boldsymbol \theta^-$ equal to $\boldsymbol \theta$ and keep it unchanged for the next $B$ updates of $\boldsymbol \theta$. Thus, the loss in \eqref{eq:target1} is modified to
\begin{align}
\left(R_n +\gamma \underset{a\in \mathcal A}{\max}\ \hat Q (S_{n+1},a; \boldsymbol \theta^-)-\hat Q(S_n, A_n;\boldsymbol \theta) \right)^2. \label{eq:target}
\end{align}
This  helps to keep the target remaining relatively stable and hence improve the convergence property of the training.

Another essential technique proposed in \cite{1186} is {\it experience replay}, which stores the {\it state-action-reward-nextState} transition $(S_n, A_n, R_n, S_{n+1})$ into a replay memory and then randomly access them later to perform the weight updates. Together with mini-batch update, i.e., multiple experiences are sampled uniformly at random from the replay memory for each update, experience replay helps to reduce the variance of the updates by avoiding highly correlated data for successive updates.

Soon after the introduction of DQN in \cite{1186}, many techniques for further performance improvement have been proposed, such as DDQN \cite{1195} to address the over-estimation bias of the maximum operation of the Q learning, DQN with prioritized experience replay (PER) \cite{1196}, DQN with the dueling network architecture \cite{1187}, and a combination of all major improvement techniques, termed {\it Rainbow}, has been introduced in \cite{1197}.


\section{Proposed Algorithms}\label{sec:ProposedAlgo}
In this section, we present our proposed DRL-based algorithms for solving the coverage-aware UAV navigation problem (P1).
\subsection{Coverage-Aware UAV Navigation as an MDP}\label{sec:UAVasMDP}
The first step to apply RL algorithms for solving a real-world problem is to reformulate it as an MDP. As MDP is defined over discrete time steps, for the UAV navigation and cell association problem $\mathrm{(P1)}$, we need to first discretize the time horizon $[0,T]$ into $N$ time steps with certain interval $\Delta_t$, where $T=N\Delta_t$. Note that $\Delta_t$ should be small enough so that within each time step, the distance between the UAV and any BS in the area of interest are approximately unchanged, or their large-scale channel gain and the BS antenna gain towards the UAV, i.e., $\beta_m(\mathbf q(t))$ and $G_m(\mathbf q(t))$ in \eqref{eq:pmt}, remain approximately constant.
As such, the UAV trajectory $\{\mathbf q(t)\}$ can be approximately represented by the sequence $\{\mathbf q_n\}_{n=1}^N$, thus \eqref{eq:DifferentialEq} and \eqref{eq:NormConstr} can be rewritten  as
\begin{align}
&\mathbf q_{n+1}=\mathbf q_n+\Delta_s \vec{\mathbf v}_n, \ \forall n, \\
& \|\vec{\mathbf v}_n\|=1, \ \forall n,
\end{align}
where $\Delta_s=V_{\max} \Delta_t$ is the UAV (maximum) displacement per time step, and $\vec{\mathbf v}_n\triangleq \vec{\mathbf v}(n\Delta_t)$ denotes the UAV flying direction at time step $n$. Furthermore, as cell association is usually based on large-scale channel gains with BSs to avoid too frequent handover, the associated cell thus remains unchanged within each time step. Thus, the association policy $b(t)$ can be represented in discrete time as $b_n$. As a result, the expected outage duration \eqref{eq:barTout} can be approximately rewritten as
\begin{align}
\bar T_{\out}(\left \{\mathbf q(t),b(t)\right\})
&\approx \Delta_t\sum_{n=1}^N P_\out(\mathbf q_n,b_n). \label{eq:barToutApprox}
\end{align}

For each time step $n$ with given UAV location $\mathbf q_n$ and cell association $b_n$, deriving a closed-form expression for $P_\out(\mathbf q_n,b_n)$ is difficult as previously mentioned in Section~\ref{sec:SystemModel}. Fortunately, this issue can be circumvented by noting the fact that the time step $\Delta_t$, which corresponds to UAV trajectory discretization based on the large-scale channel variation only, is actually in a relatively large time scale (say fractions of a second) and typically contains a large number of channel coherence blocks (say on the order of milliseconds) due to the small-scale fading. This thus provides a practical method to evaluate \eqref{eq:barToutApprox} numerically based on the raw signal measurement at the UAV. To this end, for given $\mathbf q_n$ and $b_n$ at time step $n$, we first denote the instantaneous SIR in \eqref{eq:gammat} as $\SIR(\mathbf q_n, b_n;\tilde{h})$, where $\tilde{h}$ includes all the random small-scale fading coefficients with the $M$ cells in \eqref{eq:pmt}. Further define an indicator function $I(\cdot)$ as
\begin{align}\label{eq:indicator}
I(\mathbf q, b; \tilde h)=\begin{cases}
1, \ \text{if } \SIR(\mathbf q,b;\tilde h)< \gamma_{\thr} \\
0, \ \text{otherwise}.
\end{cases}
\end{align}
Then we have
\begin{align}
P_\out(\mathbf q_n,b_n)&=\Pr\left(\SIR(\mathbf q_n,b_n;\tilde h)< \gamma_{\thr} \right)\\
&= \mathbb{E}_{\tilde h} \left[I(\mathbf q_n,b_n;\tilde h)\right]. \label{eq:barToutApprox2}
\end{align}

  Assume that within each time step $n$, the UAV performs $J$ SIR measurements for each of the $M$ cell associations, which could be achieved by leveraging the existing soft handover mechanisms with continuous RSRP and RSRQ reports. Denote the $j$th SIR measurement of time step $n$ with cell association $b_n\in\{1,\cdots, M\}$ as $\SIR(\mathbf q_n, b_n; \tilde h[n,j])$, where $\tilde h[n,j]$ denotes the realization of the small-scale fading. The corresponding outage indicator value, denoted as $I(\mathbf q_n,b_n;\tilde h[n,j])$,  can be obtained accordingly based on \eqref{eq:indicator}. Then the empirical outage probability can be obtained as
\begin{align}
\hat P_\out(\mathbf q_n,b_n)\triangleq \frac{1}{J}\sum_{j=1}^J I(\mathbf q_n,b_n;\tilde h[n,j]).
\end{align}
Then by applying the Law of Large Numbers to \eqref{eq:barToutApprox2}, we have $\lim_{J\rightarrow \infty}\hat P_\out(\mathbf q_n,b_n)=P_\out(\mathbf q_n,b_n)$.
Therefore, as long as the UAV performs signal measurements sufficiently frequently so that $J\gg 1$, $P_\out(\mathbf q_n,b_n)$ in \eqref{eq:barToutApprox} can be evaluated by its empirical value $\hat P_\out(\mathbf q_n,b_n)$.

Based on the above discussion, $\mathrm{(P1)}$ can be approximated as
\begin{align}
\mathrm{(P2):} \underset{N, \{\mathbf q_n, \vec{\mathbf v}_n,b_n\}}{\max} \vspace{-0.1in}  \ & -N - \mu \sum_{n=1}^N \hat P_\out(\mathbf q_n,b_n) \notag \\
\mathrm{s.t.} \  & \mathbf q_{n+1}=\mathbf q_n+\Delta_s \vec{\mathbf v}_n, \ \forall n, \label{eq:stateTrans}\\
& \|\vec{\mathbf v}_n\|=1, \ \forall n, \\
& \mathbf q_0=\mathbf q_I, \mathbf q_N=\mathbf q_F,\\
& \mathbf q_L \preceq \mathbf q_n \preceq \mathbf q_U, \forall n, \label{eq:lstConstr}\\
& b_n\in \{1,\cdots, M\}.
\end{align}

Note that we have dropped the constant coefficient $\Delta_t$ in the objective of $\mathrm{(P2)}$. Obviously, the optimal cell association policy to $\mathrm{(P2)}$ can be easily obtained as $b^*_n=\mathrm{arg}\underset{b\in \{1,\cdots, M\}}{\min} \hat P_\out(\mathbf q_n,b)$. With a slight abuse of notation, let
\begin{align}
\hat P_\out(\mathbf q_n)=\underset{b\in \{1,\cdots, M\}}{\min} \hat P_\out(\mathbf q_n,b),\label{eq:hatPout}
\end{align}
then $\mathrm{(P2)}$ reduces to
\begin{align}
\mathrm{(P3):} \underset{N, \{\mathbf q_n, \vec{\mathbf v}_n\}}{\max} \vspace{-0.1in}  \ & -N - \mu \sum_{n=1}^N \hat P_\out(\mathbf q_n) \notag \\
\mathrm{s.t.} \  & \eqref{eq:stateTrans}-\eqref{eq:lstConstr}. \notag
\end{align}

As such, a straightforward mapping of $\mathrm{(P3)}$ to an MDP $<\mathcal S, \mathcal A, \mathcal P, \mathcal R>$ is obtained as follows
\begin{itemize}
\item $\mathcal S$: the state corresponds to the UAV location $\mathbf q_n$, and the state space constitutes all possible UAV locations within the feasible region, i.e., $\mathcal S=\{\mathbf q: \mathbf q_L \preceq \mathbf q \preceq \mathbf q_U\}$.
\item $\mathcal A$: the action space corresponds to the UAV flying direction, i.e., $\mathcal A=\{\vec{\mathbf v}: \|\vec{\mathbf v}\|=1\}$.
\item $\mathcal P$: the state transition probability is deterministic governed by \eqref{eq:stateTrans}. 
\item $\mathcal R$: the reward $R(\mathbf q)=-1-\mu \hat P_\out(\mathbf q)$, so that every time step that the UAV takes would incur a penalty of $1$, and an additional penalty with weight $\mu$ if the UAV enters into a location with certain outage probability. This will thus encourage the UAV  to reach the destination $\mathbf q_F$ as soon as possible, while avoiding the weak coverage regions of the cellular network.
\end{itemize}

With the above MDP formulation, it is observed that the objective function of $\mathrm{(P3)}$ corresponds to the un-discounted  (i.e., $\gamma=1$) accumulated rewards over one {\it episode} up to time step $N$, i.e., $G_1=\sum_{n=1}^N R_n$. This corresponds to one particular form of MDP, namely the {\it episodic tasks} \cite{1084}, which are tasks containing a special state called the {\it terminal state} that separates the agent-environment interactions into {\it episodes}. For episodic tasks, each episode ends with the terminal state, followed by a random reset to the starting state to start the new episode. For the UAV navigation problem $\mathrm{(P3)}$, the terminal state corresponds to $\mathbf q_F$ and the start state is $\mathbf q_I$.
After being formulated as an MDP, $\mathrm{(P3)}$ can be solved by applying various RL algorithms. In the following, we first introduce a solution based on the state-of-the-art dueling DDQN with multi-step learning, and then propose the SNARM framework to improve the learning efficiency.

\subsection{Dueling DDQN Multi-Step Learning for UAV Navigation}\label{sec:TD}
Both the state and action spaces for the MDP defined in  Section~\ref{sec:UAVasMDP} are continuous. The most straightforward approach for handling continuous state-action spaces is to discretize them. However, a naive discretization of both state and action spaces either leads to poor performance due to the discretization errors and/or results in a prohibitively large number of state-action pairs due to the curse of dimensionality \cite{1084}. In this paper, we only discretize the action space $\mathcal A$ while keeping the state space $\mathcal S$ as continuous, and use the powerful ANN to approximate the action-value function.  By uniformly discretizing  the action space $\mathcal A$ (i.e., the flying directions) into $K$ values, we have $\hat {\mathcal A}=\{\vec{\mathbf v}^{(1)},\cdots, \vec{\mathbf v}^{(K)}\}$. 

\begin{figure}
\centering
\includegraphics[width=0.65\textwidth]{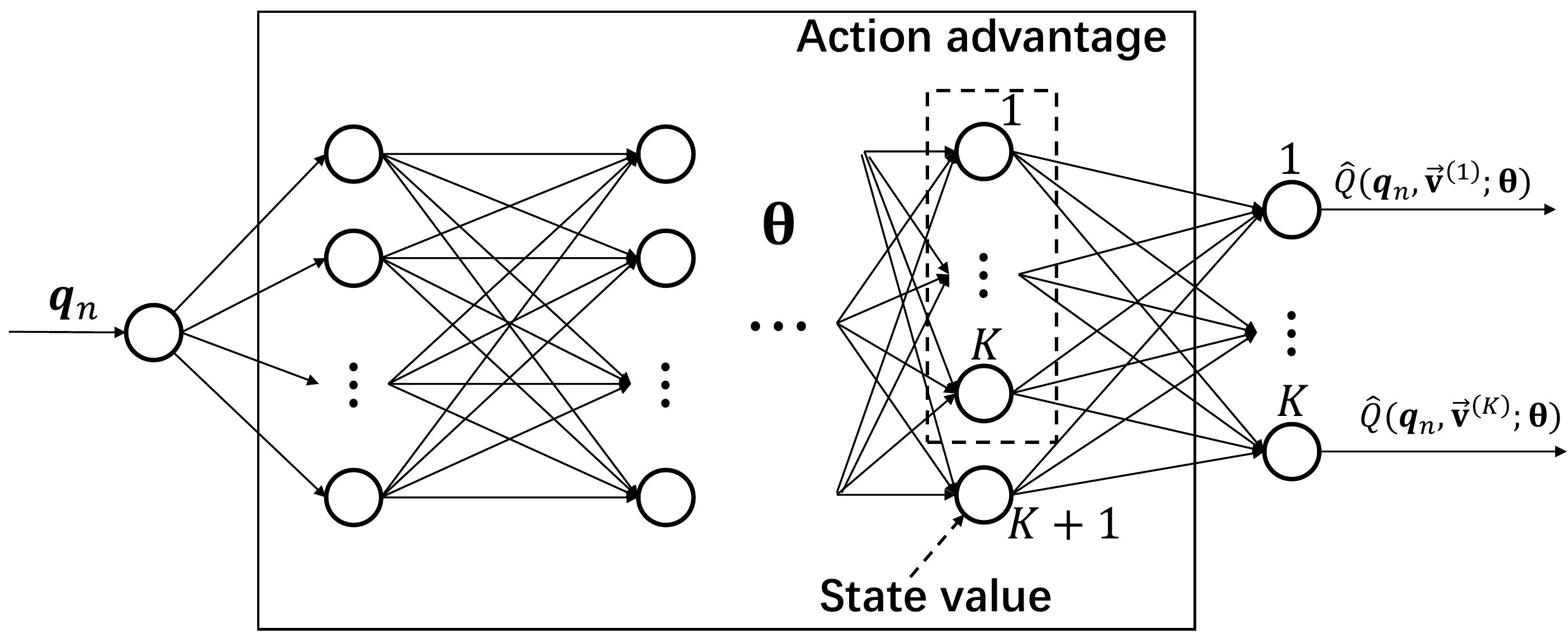}
\caption{Dueling DQN for coverage-aware UAV navigation.\vspace{-3ex}}
\label{F:DQN}
\end{figure}

With the above action space discretization, the action-value function $Q(\mathbf q, \vec{\mathbf v})$ contains the continuous state input $\mathbf q\in \mathcal S$ and the discrete action input $\vec{\mathbf v}\in \hat {\mathcal A}$. We use the state-of-the-art DQN with the dueling network architecture \cite{1187} to approximate $Q(\mathbf q, \vec{\mathbf v})$, as illustrated in Fig.~\ref{F:DQN}. At each time step $n$, the input of the network corresponds to the state (or the UAV location) $\mathbf q_n\in \mathcal S$, and it has $K$ outputs, each corresponding to one action in $\hat {\mathcal A}$. Note that the distinguishing feature of dueling DQN is to first separately estimate the state values and the state-dependent {\it action advantages}, and then combine them in a smart way via an aggregating layer to give an estimate of the action value function Q  \cite{1187}. Compared to the standard DQN, dueling DQN is able to learn the state-value function more efficiently, and is also more robust to the approximation errors when the action-value gaps between different actions of the same state are small \cite{1187}. The coefficients of the dueling DQN are denoted as $\boldsymbol \theta$, which is trained so that the output $\hat Q(\mathbf q,\vec{\mathbf v};\boldsymbol \theta)$ gives good approximations to the true action-value function $Q(\mathbf q,\vec{\mathbf v})$.

To train the dueling DQN, besides the standard techniques of {\it experience replay} and {\it target network} as discussed in Section~\ref{sec:DRL}, we also apply the multi-step DDQN learning techniques \cite{1195}. Specifically, define the truncated $N_1$-step return from a given state $\mathbf q_n$ as
\begin{align}
R_{n:n+N_1}=\sum_{i=0}^{N_1-1}\gamma^i R_{n+i+1}.\label{eq:RMultistep}
\end{align}
Then a multi-step DQN learning is to minimize the loss given by \cite{1084}
\begin{align}
\left(R_{n:n+N_1}+\gamma^{N_1}\underset{\vec{\mathbf v}'\in \hat {\mathcal A}}{\max}\ \hat Q(\mathbf q_{n+N_1},\vec{\mathbf v}'; \boldsymbol \theta^-) - \hat Q(\mathbf q_n,\vec{\mathbf v}_n; \boldsymbol \theta) \right)^2. \label{eq:multiStepLearning}
\end{align}
It is known that multi-step learning with appropriately chosen $N_1$ usually leads to faster learning \cite{1084}.

 Furthermore, double Q learning has been widely used to address the overestimation bias of the maximum operation in \eqref{eq:multiStepLearning}, via using two separate networks for the action selection and value evaluation of bootstrapping. Specifically, with DDQN, the loss of the multi-step learning in \eqref{eq:multiStepLearning} is changed to
\begin{align}
\left(R_{n:n+N_1}+\gamma^{N_1}\hat Q(\mathbf q_{n+N_1},\vec{\mathbf v}^*; \boldsymbol \theta^-) - \hat Q(\mathbf q_n,\vec{\mathbf v}_n; \boldsymbol \theta) \right)^2, \label{eq:multiStepLearningDDQN}
\end{align}
where
\begin{align}
\vec{\mathbf v}^*=\underset{\vec{\mathbf v}'\in \hat{\mathcal A}}{\mathrm{arg}\max}\ \hat Q(\mathbf q_{n+N_1},\vec{\mathbf v}'; \boldsymbol \theta).\label{eq:actionSel}
\end{align}
Note that we have used the target network with coefficients $\boldsymbol \theta^-$ to evaluate the bootstraping action in \eqref{eq:multiStepLearningDDQN} while using the Q network with  coefficients $\boldsymbol \theta$ for action selection in \eqref{eq:actionSel}.

The proposed algorithm for coverage-aware UAV navigation with dueling DDQN multi-step learning is summarized in Algorithm~\ref{Algo:PathLearning}. Note that the main steps of Algorithm~\ref{Algo:PathLearning} follow from the classic DQN algorithms in \cite{1186}, except the following modifications. First, the more advanced dueling DQN architecture is used, instead of DQN. Second, to enable multi-step learning, a sliding window queue with capacity $N_1$ as in steps 6 and 11 is used to store the $N_1$ latest transitions. Therefore, at the current time step $n$, the UAV can calculate the accumulated return from the previous $N_1$ time steps, as in step 12 of Algorithm~\ref{Algo:PathLearning}. Third, a double DQN learning is used in step 14. 

\begin{algorithm}
\caption{Dueling DDQN Multi-Step Learning for Coverage-Aware UAV Navigation.}\label{Algo:PathLearning}
\begin{algorithmic}[1]
\STATE {\bf Initialize:} the maximum number of episodes $\bar{N}_{\mathrm{epi}}$,  maximum number of steps per episode $\bar{N}_{\mathrm{step}}$, initial exploration  $\epsilon_0$, decaying rate $\alpha$, set $\epsilon\leftarrow \epsilon_0$
\STATE {\bf Initialize:} reaching-destination reward $R_{\mathrm{des}}$, outbound penalty $P_{\mathrm{ob}}$, outage penalty weight $\mu$
\STATE {\bf Initialize:} the replay memory queue $D$ with capacity $C$
\STATE {\bf Initialize:} the dueling DQN network with coefficients $\boldsymbol \theta$, the target network with coefficients $\boldsymbol \theta^-=\boldsymbol \theta$ \FOR{$n_{\mathrm{epi}}=1,\cdots, \bar{N}_{\mathrm{epi}}$}
\STATE Initialize a sliding window queue $W$ with capacity $N_1$
\STATE Randomly initialize the state $\mathbf q_I\in \mathcal S$, set the time step $n\leftarrow 0$ and $\mathbf q_0 \leftarrow  \mathbf q_I$.
\REPEAT
\STATE Choose action $\vec{\mathbf v}_n$ from $\hat {\mathcal A}$ based on the $\epsilon$-greedy policy, 
i.e., $\vec{\mathbf v}_n=\vec{\mathbf v}^{(k^*)}$, where
\begin{equation}\notag 
\small
\begin{aligned}
k^*=\begin{cases}
\mathrm{randi}(K), &  \text{ with prob. }  \epsilon, \\
\underset{k=1,\cdots, K}{\mathrm{argmax}} \hat Q(\mathbf q_n, \vec{\mathbf v}^{(k)};\boldsymbol \theta), & \text{ with prob. } 1-\epsilon.
\end{cases}
\end{aligned}
\end{equation}
\STATE Execute action $\vec{\mathbf v}_n$ and observe the next state $\mathbf q_{n+1}$, measure the signals and obtain the empirical outage probability $\hat P_{\out}(\mathbf q_{n+1})$ as in \eqref{eq:hatPout}, and set the reward as $R_n=-1-\mu \hat P_{\out}(\mathbf q_{n+1})$
\STATE Store transition $(\mathbf q_n, \vec{\mathbf v}_n, R_n, \mathbf q_{n+1})$ in the sliding window queue $W$
\STATE If $n\geq N_1$, calculate the $N_1$-step accumulated return $R_{(n-N_1):n}$ based on \eqref{eq:RMultistep} using the stored transitions in $W$, and store the $N_1$-step transition $(\mathbf q_{n-N_1}, \vec{\mathbf v}_{n-N_1}, R_{(n-N_1):n}, \mathbf q_n)$ in the replay memory $D$
\STATE Sample random minibatch of $N_1$-step transition $(\mathbf q_j, \vec{\mathbf v}_j, R_{j:j+N_1}, \mathbf q_{j+N_1})$ from $D$
\STATE Set
\begin{equation}\notag
\small
\hspace{-4ex}
\begin{aligned}
y_j=\begin{cases}
R_{j:j+N_1}+R_{\mathrm{des}}, & \text{ if }  \mathbf q_{j+N_1}=\mathbf q_F \\
R_{j:j+N_1}-P_{\mathrm{ob}}, & \text{ if } \mathbf q_{j+N_1} \notin \mathcal{S} \\
R_{j:j+N_1}+ \gamma^{N_1}\hat Q(\mathbf q_{j+N_1},\vec{\mathbf v}^*; \boldsymbol \theta^-), & \text{ otherwise },
\end{cases}
\end{aligned}
\end{equation}
where {\small $\vec{\mathbf v}^*=\underset{\vec{\mathbf v}'\in \hat{\mathcal A}}{\mathrm{arg}\max}\ \hat Q(\mathbf q_{j+N_1},\vec{\mathbf v}'; \boldsymbol \theta)$}.
\STATE Perform a gradient descent step on $\big(y_j - \hat Q(\mathbf q_j,\vec{\mathbf v}_j;\boldsymbol \theta)\big)^2$ with respect to network parameters $\boldsymbol \theta$
\STATE Update $n\leftarrow n+1$, $\epsilon\leftarrow \epsilon\alpha$
\UNTIL $\mathbf q_n= \mathbf q_F$ or $\mathbf q_n\notin \mathcal{S}$, or $n=\bar{N}_{\mathrm{step}}$.
\STATE After every $B$ episodes, set the target network parameters $\boldsymbol \theta^-\leftarrow \boldsymbol \theta$
\ENDFOR
\end{algorithmic}
\end{algorithm}

While theoretically, the convergence of Algorithm~\ref{Algo:PathLearning} is guaranteed \cite{1084}, in practice, a random initialization of the dueling DQN may require a prohibitively large number of time steps for the UAV to reach the destination $\mathbf q_F$. Intuitively, $\boldsymbol \theta$ should be initialized in a way such that in the early episodes when the UAV has no or limited knowledge about the radio environment, it should be encouraged to select actions for the shortest path flying. Thus, we propose the {\it distance-based} initialization for Algorithm~\ref{Algo:PathLearning}, with $\boldsymbol \theta$ initialized so that $\hat Q(\mathbf q, \vec{\mathbf v}; \boldsymbol \theta)$ approximates $-\|\mathbf q'-\mathbf q_F\|$, where $\mathbf q'$ is the next state with current state $\mathbf q$ and action $\vec{\mathbf v}$. Note that as the distance from any location $\mathcal S$ to the destination $\mathbf q_F$ is known {\it a priori}, the above initialization can be performed by sampling some locations in $\mathcal S$ without requiring any interaction with the environment. With such an initialization, it can be obtained from step 9 of Algorithm~\ref{Algo:PathLearning}  that the UAV will choose the shortest-path action in the first episode with probability $1-\epsilon$, and random exploration with probability $\epsilon$. As the UAV accumulates more experience after some episodes so that the dueling DQN has been substantially updated, the $\epsilon$-greedy policy in step 9 would generate UAV flying path that achieves a balance between minimizing the flying distance and avoiding the weak coverage region, as desired for problem $\mathrm{(P3)}$.

\subsection{Simultaneous Navigation and Radio Mapping}
The dueling DDQN multi-step learning proposed in Algorithm~\ref{Algo:PathLearning} is a {\it model-free} Q learning algorithm, where all data used to update the estimation of the action-value function is obtained via the direct interaction of the UAV with the environment, as in step 10 of Algorithm~\ref{Algo:PathLearning}. While model-free RL requires no prior knowledge about the environment nor intending to estimate it, it usually leads to slow learning process and requires a large number of agent-environment interactions, which is typically costly or even risky to obtain. For instance, for the coverage-aware UAV navigation problem considered in this paper,  each step 10 in Algorithm~\ref{Algo:PathLearning} actually requires to fly the UAV and perform signal measurements at designated locations to get the empirical outage probability so as to obtain the reward value. Extensive UAV flying increases the safety risk due to e.g., mechanical fatigue. To overcome the above issues, in this subsection, we propose a novel SNARM framework for cellular-connected UAVs by utilizing the dueling DDQN multi-step learning algorithm with {\it model learning} \cite{1084}.

For RL with model learning, each real experience obtained by the UAV-environment interaction can actually be used for at least two purposes: for {\it direct RL} as in Algorithm~\ref{Algo:PathLearning} and for {\it model learning} so as to gain useful information about the environment. A closer look at the MDP corresponding to problem $\mathrm{(P3)}$ would reveal that the essential information needed for coverage-aware navigation is the reward function $R(\mathbf q)$, which depends on the outage probability $P_\out(\mathbf q)$ for any location $\mathbf q$ visited by the UAV. Therefore, model learning for $\mathrm{(P3)}$ corresponds to learning the {\it radio map} of the airspace where the UAVs are operating, which specifies the cellular outage probability $P_{\out}(\mathbf q)$, $\forall \mathbf q\in \mathcal S$. Thus, the actual measurement of the outage probability $\hat P_\out(\mathbf q)$ at location $\mathbf q$ can be simultaneously used for navigation, i.e., determining the UAV flying direction $\vec{\mathbf v}$ for the next time step, and also for radio mapping, i.e., helping to predict the outage probability for locations that the UAV has not visited yet. This leads to our proposed framework of SNARM.

\begin{figure}
\centering
\includegraphics[width=0.65\textwidth]{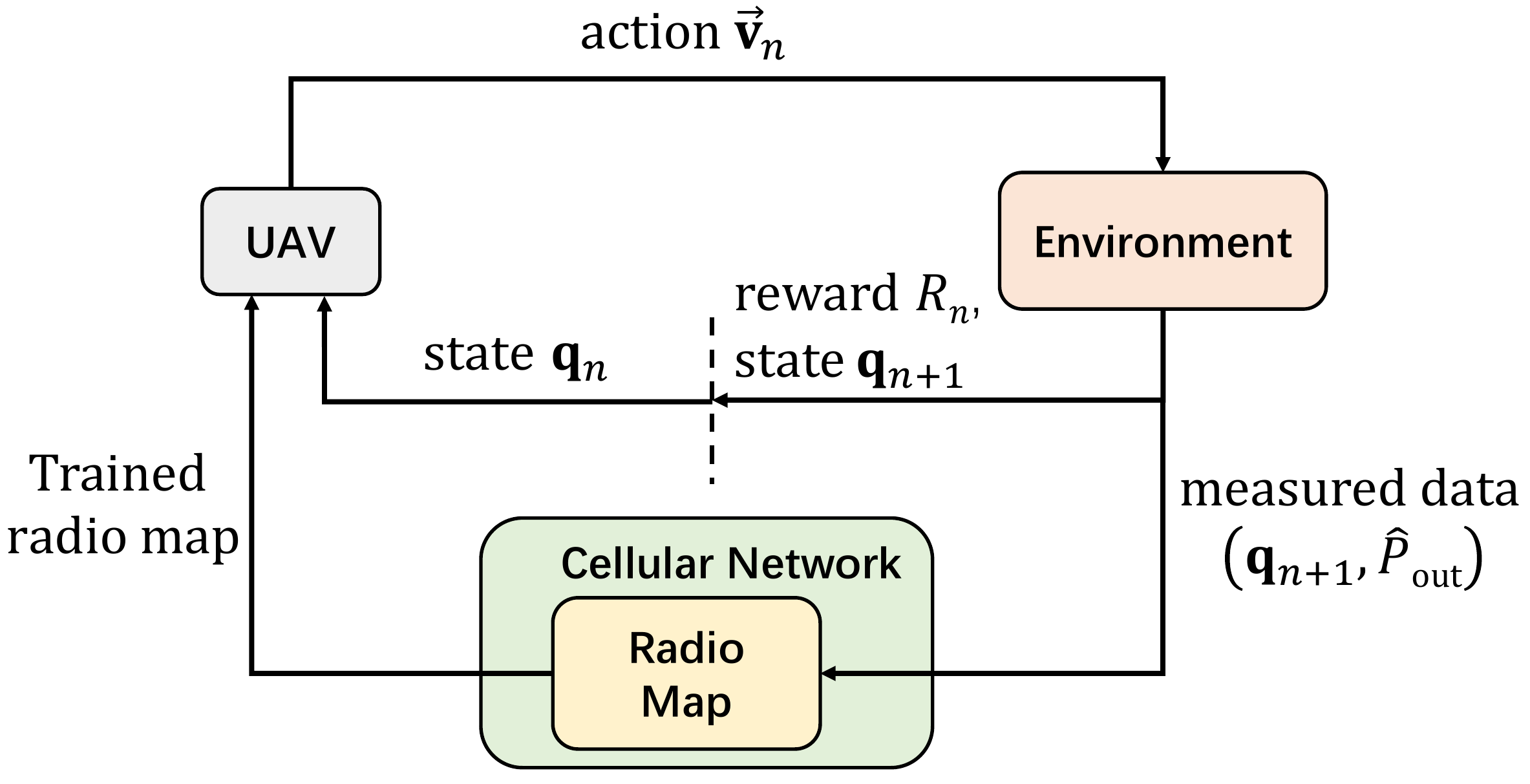}
\caption{SNARM for cellular-connected UAV.\vspace{-3ex}}
\label{F:SNARM}
\end{figure}

A high-level overview of SNARM is illustrated in Fig.~\ref{F:SNARM}. Different from the standard RL setup, with SNARM for cellular-connected UAV, the cellular network (say, a selected macro-cell in the area of interest) maintains a radio map $\mathcal M=\{P_\out(\mathbf q), \forall \mathbf q\in \mathcal S\}$ containing an estimation of the cellular outage probability for all locations in $\mathcal S$. Note that $\mathcal M$ might be highly inaccurate initially, but can be continuously improved as more real experience is accumulated. For any cellular-connected UAV flying in the considered airspace $\mathcal S$, it may either download the radio map $\mathcal M$ from the network off-line before the mission starts, or continuously receive the latest update of $\mathcal M$ in real time. At each time step $n$ as the UAV performs its mission, it observes its current location $\mathbf q_n$,  together with the radio map $\mathcal M$, then determines its flying direction $\vec{\mathbf v}_n$ for the next step. After executing the action and reaching the new location $\mathbf q_{n+1}$, it measures the empirical outage probability $\hat P_\out(\mathbf q_{n+1})$, and receives the corresponding reward $R_n$. The obtained empirical outage probability $\hat P_\out(\mathbf q_{n+1})$ based on signal measurement can be used as the new data input to improve the radio map estimation. In the following, we explain the radio map update and UAV action selection of SNARM, respectively.

 With any finite measurements $\{\mathbf q_n, \hat P_\out(\mathbf q_n)\}$, predicting the outage probability $P_\out(\mathbf q), \forall \mathbf q\in \mathcal S$ is essentially a {\it supervised learning} problem. In this paper, we use a feedforward fully-connected ANN with parameters $\boldsymbol \xi$ to represent the radio map, i.e., $\boldsymbol \xi$ is trained so that $P_\out(\mathbf q)\approx \hat P_\out(\mathbf q; \boldsymbol \xi)$, $\forall \mathbf q\in \mathcal S$. Note that compared to the standard supervised learning problem, the radio mapping problem has two distinct characteristics. First, the data measurement $\{\mathbf q_n, \hat P(\mathbf q_n)\}$ only arrives incrementally as the UAV flies to new locations, instead of all made available at the beginning of the training. Second, as the UAV performs signal measurements as it flies, those data obtained at consecutive time steps are typically correlated, which is undesirable for supervised training. To tackle these two issues, similar to Algorithm~\ref{Algo:PathLearning}, we use a database (replay memory) to store the measured data $\{\mathbf q_n, \hat P(\mathbf q_n)\}$, and at each training step, a minibatch is sampled at random from the database to update the network parameter $\boldsymbol \xi$. 

 On the other hand, the problem for UAV  action selection with the assistance of a radio map $\mathcal M$ is known as {\it indirect RL}, for which various algorithms have been proposed \cite{1084}. Different from that for direct RL in Algorithm~\ref{Algo:PathLearning}, with the radio map $\mathcal M$ in SNARM, the UAV is able to predict the estimated return for each trajectory it would take, without having to actually fly along it. As a result, we may generate as much simulated experience as we wish based on $\mathcal M$, which, together with the real experience, can be used to update the UAV policy based on any RL algorithm, such as Algorithm~\ref{Algo:PathLearning}. However, as the radio map might be inaccurate at the initial stages, over-relying on the simulated experience may result in poor performance due to the model error. Therefore, we need to seamlessly integrate the learning with simulated experience based on the radio map and that with the real experience. A simple but effective architecture achieving this goal is known as {\it Dyna} \cite{1084}, where for each RL update based on the real experience, we have several updates based on the simulated experience. The proposed SNARM based on Dyna dueling DDQN with multi-step learning is summarized in Algorithm~\ref{Algo:SNARM}.

\begin{algorithm}
\caption{Dyna dueling DDQN with multi-step learning for SNARM.}\label{Algo:SNARM}
\begin{algorithmic}[1]
\STATE {Download the radio map $\hat P_\out(\mathbf q; \boldsymbol \xi)$ from the cellular network, and denote the existing database containing all measurements $\{\mathbf q, \hat P_\out(\mathbf q)\}$ as $E$}
\STATE Steps 1--7 of Algorithm~\ref{Algo:PathLearning}
\STATE Randomly initialize the simulated starting state $\tilde {\mathbf q}_I\in \mathcal S$, set the simulated time step $\tilde n\leftarrow 0$ and $\tilde {\mathbf q}_0\leftarrow \tilde {\mathbf q}_I$
\STATE Steps 8--16 of Algorithm~\ref{Algo:PathLearning}
\STATE Add the measured data $(\mathbf q_{n+1}, \hat P_\out(\mathbf q_{n+1}))$ to $E$
\STATE Sample random minibatch from $E$ and update the  network parameter $\boldsymbol \xi$ for radio map
\FOR{$i=1,\cdots, \tilde{N}$}
\STATE Choose action for the simulated state $\tilde {\mathbf q}_{\tilde n}$ based on the $\epsilon$-greedy policy similar as step 9 of Algorithm~\ref{Algo:PathLearning}
\STATE Obtain the simulated next state $\tilde{\mathbf q}_{\tilde n+1}$, predict the outage probability $\hat P_{\out}(\tilde{\mathbf q}_{\tilde n+1}; \boldsymbol \xi)$ based on the radio map $\mathcal M$ and the corresponding reward $\tilde R_{\tilde n}=-1-\mu \hat P_{\out}(\tilde{\mathbf q}_{\tilde n+1}; \boldsymbol \xi)$.
\STATE Perform similar operations as steps 11--15 of Algorithm~\ref{Algo:PathLearning} for the simulated experience
\STATE Update the simulated time step $\tilde n\leftarrow \tilde n+1$
\STATE If $\tilde{\mathbf q}_{\tilde n}=\mathbf q_F$ or $\tilde{\mathbf q}_{\tilde n}\notin \mathcal S$ or $\tilde n=\bar N_{\mathrm{step}}$, reset the simulated trajectory as in step 3
\ENDFOR
\STATE Steps 17--19 of Algorithm~\ref{Algo:PathLearning}.
\end{algorithmic}
\end{algorithm}

Obviously, Algorithm~\ref{Algo:SNARM} expands Algorithm~\ref{Algo:PathLearning} by inserting the radio map learning operations and the double DQN update based on simulated UAV experience. Note that the simulated trajectory is initialized independently from the real trajectory (see steps 3 and 12 of Algorithm~\ref{Algo:SNARM}), and it is used $\tilde N$ times more frequently than the real trajectory, i.e., for each actual step the UAV takes in real experience, $\tilde N$ steps will be taken in the simulated trajectory, as in steps 7 to 13. Note that in general, $\tilde N$ should be increasing with the number of episodes, since as more real experience is accumulated, the constructed radio map becomes more accurate, which makes the predicted returns of the simulated trajectories more reliable. Regardless of simulated or real experience, the same action-value update based on dueling DDQN multi-step learning is performed. Besides, the radio map is updated in step 6 based on the actual measurement data, and used in step 9 to generate simulated UAV experience.


\section{Numerical Results}\label{sec:Numerical}
\begin{figure}
\centering
\begin{subfigure}[b]{0.47\textwidth}
\centering
\includegraphics[width=\textwidth]{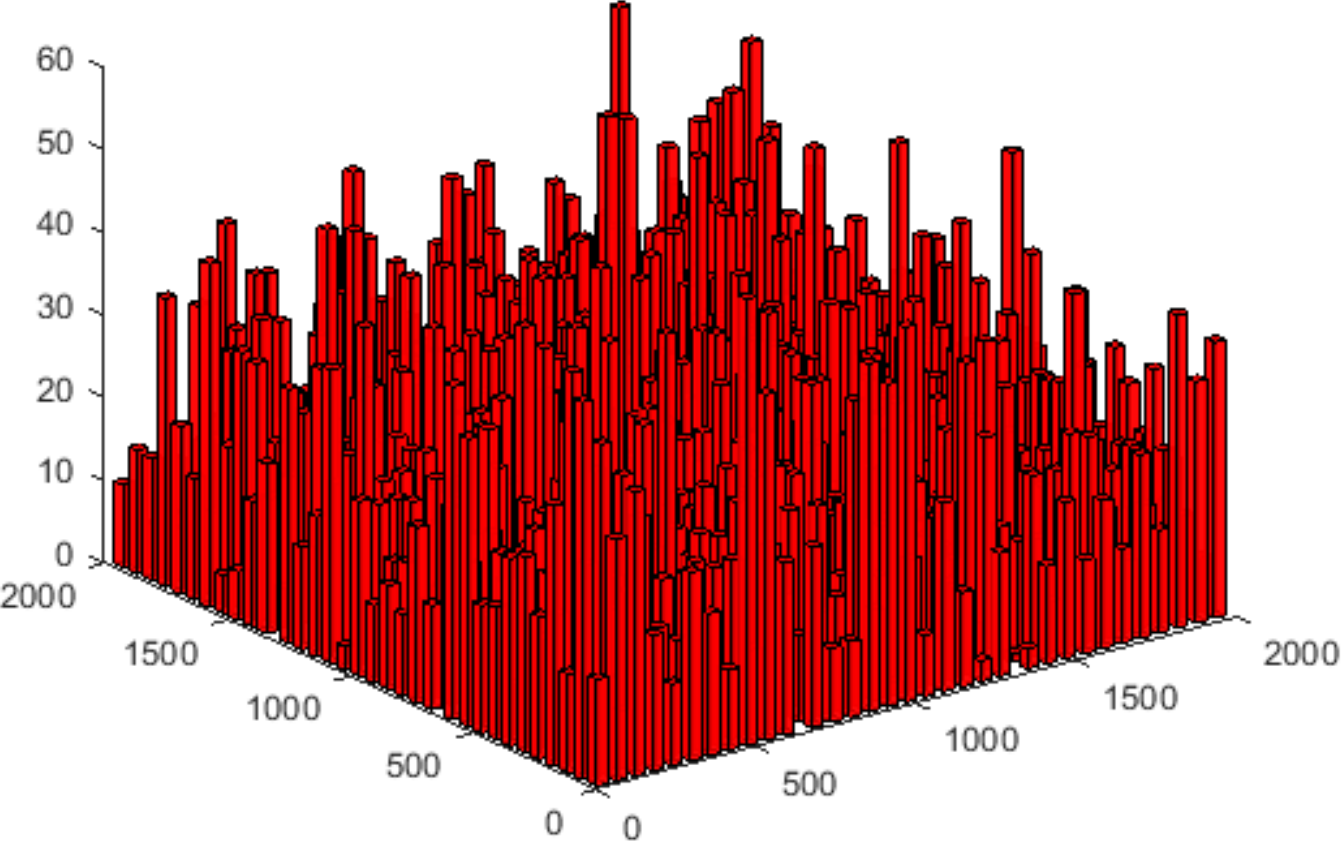}
\caption{3D view}
\end{subfigure}
\hfill
\begin{subfigure}[b]{0.47\textwidth}
\centering
\includegraphics[width=\textwidth]{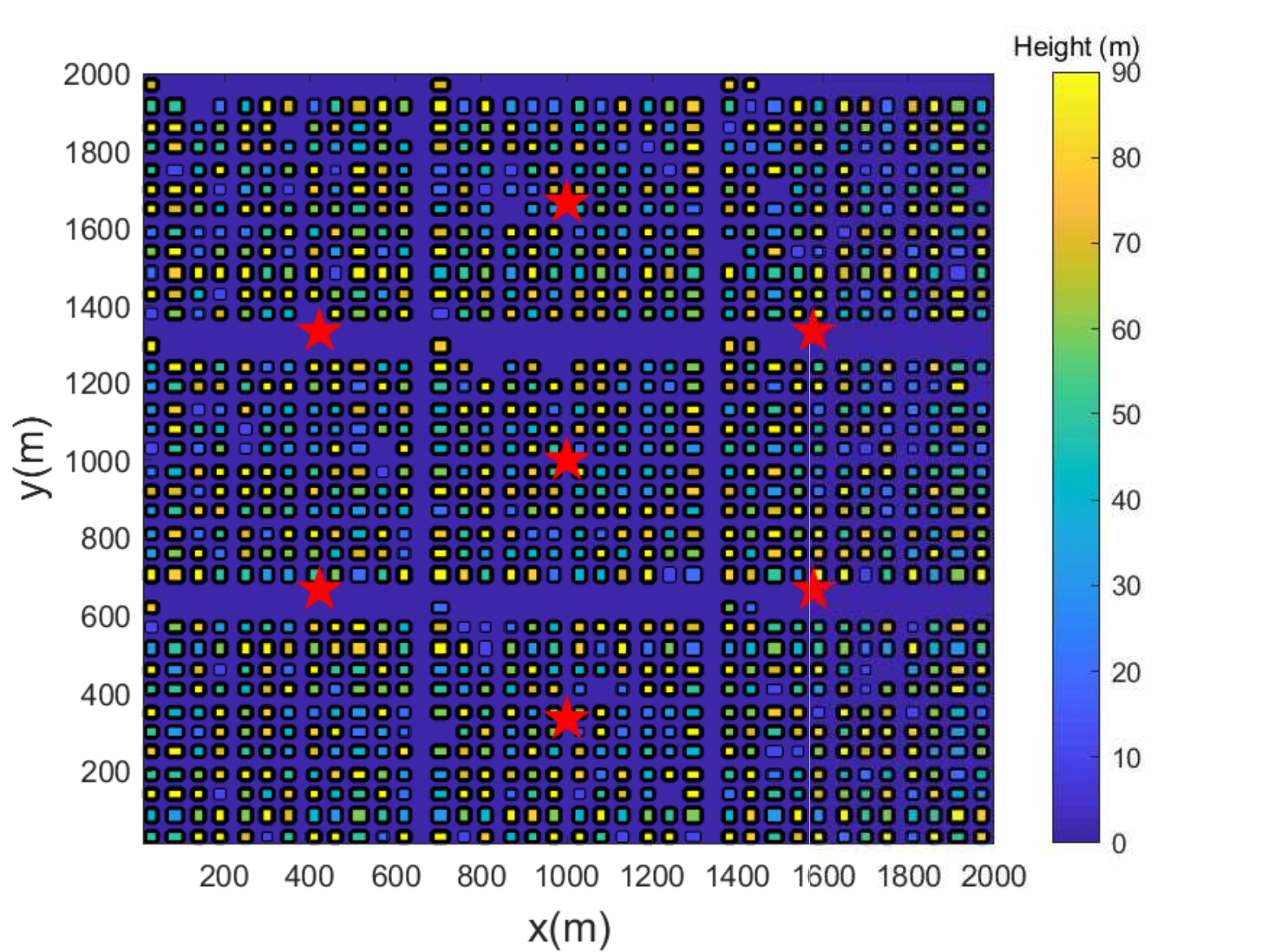}
\caption{Top view}
\end{subfigure}
\caption{The building locations and heights in 3D and 2D views. The BS locations are marked by red stars, where each BS site contains 3 cells.\vspace{-2ex}}
\label{F:building3D}
\end{figure}

In this section, numerical results are provided to evaluate the performance of the proposed algorithms. 
As shown in Fig.~\ref{F:building3D}, we consider an urban area of size 2 km $\times$ 2 km with high-rise buildings, which corresponds to the most challenging environment for coverage-aware UAV navigation, since the LoS/NLoS links and the received signal strength may alter frequently as the UAV flies (see Fig.~\ref{F:SystemModel}). To accurately simulate the BS-UAV channels in the given environment, we first generate the building locations and heights  based on one realization of the statistical model suggested by International Telecommunication Union (ITU) \cite{1094}, which involves three parameters, namely, $\alpha_{\mathrm {bd}}$: the ratio of land area covered by buildings to the total land area; $\beta_{\mathrm{bd}}$: the mean number of buildings per unit area; and $\gamma_{\mathrm{bd}}$: a variable determining the building height distribution, which is  modelled as Rayleigh distribution with mean value $\sigma_{\mathrm{bd}}>0$. Note that such statistical building model has been widely used to obtain the LoS probability for aerial-ground links \cite{642}, which, however, only reflects the average characteristics over a large number of geographic areas of similar type. While for each local area with given building locations and heights, the presence/absence of LoS links with cellular BSs can be exactly determined by checking whether the line connecting the BS and UAV is blocked or not by any building. Fig.~\ref{F:building3D} shows the 3D and 2D views of one particular realization of the building locations and heights with $\alpha_{\mathrm {bd}}=0.3$, $\beta_{\mathrm{bd}}=300$ buildings/km$^2$, and $\sigma_{\mathrm{bd}}=50$ m. For convenience, the building height is clipped to not exceed $90$ m.

We assume that the considered area has 7 cellular BS sites with locations marked by red stars in Fig.~\ref{F:building3D}, and the BS antenna height is $25$ m \cite{1012}. With the standard sectorization technique, each BS site contains 3 sectors/cells. Thus, the total number of cells is $M=21$. The BS antenna model follows the 3GPP specification \cite{1024}, where an 8-element  uniform linear array (ULA) is placed vertically. Each array element itself is directional, with half-power beamwidths  along the vertical and horizontal dimensions both equal to $65^\circ$. Furthermore, pre-determined phase shifts are applied to the vertical antenna array so that its main lobe is electrically downtilted towards the ground by $10^\circ$. This leads to the directional antenna array with fixed 3D  radiation pattern, as shown in Fig.~4 of \cite{1095}. To simulate the signal strength received by the UAV from each cell,  we firstly determine whether there exists an LoS link between the UAV and each BS at the given UAV location based on the building realization,  and then generate the BS-UAV path loss using the 3GPP  model for urban Macro (UMa) given in Table B-2 of \cite{1012}. The small-scale fading coefficient is then added assuming Rayleigh fading for the NLoS case and Rician fading with 15 dB Rician factor for the LoS case. It is worth mentioning that for a given environment, the aforementioned LoS condition, antenna gain, path loss and small-scale fading with each BS are all dependent on the UAV location in a rather sophisticated manner. This makes them only suitable to generate the numerical values with given UAV location, instead of being directly used for UAV path planning with the conventional optimization technique as for solving $\mathrm{(P1)}$, even with accurate channel modelling and perfect global information. This thus justifies the use of RL techniques for UAV path design as pursued in this paper.

%

For both Algorithms~\ref{Algo:PathLearning} and ~\ref{Algo:SNARM}, the dueling DQN adopted is a fully connected feedforward ANN consisting of 5 hidden layers. The numbers of neurons of the first 4 hidden layers are 512, 256, 128, and 128, respectively. The last hidden layer is the dueling layer with $K+1$ neurons, with one neuron corresponding to the estimation of the state-value and the other $K$ neurons corresponding to the {\it action advantages} for the $K$ actions, i.e., the difference between the action values and the state value of each state. The outputs of these $K+1$ neurons are then aggregated in the output layer to obtain the estimation of the $K$ action values \cite{1187}, as shown in Fig.~\ref{F:DQN}. For SNARM in Algorithm~\ref{Algo:SNARM}, the ANN for radio mapping also has 5 hidden layers, with the corresponding numbers of neurons given by 512, 256, 128, 64, and 32, respectively. The activation for all hidden layers is by {\it ReLu} \cite{1199}, and the ANN is trained with {\it Adam} optimizer to minimize the mean square error (MSE) loss,   using Python with TensorFlow and Keras.\footnote{https://keras.io/} For ease of illustration, we assume that the UAV's flying altitude is fixed as $H=100$ m, and the number of actions (or flying directions) at each time step is $K=4$. All UAV missions have a common destination location $\mathbf q_F=[1400, 1600, 100]^T$ m, while their initial locations $\mathbf q_I$ are randomly generated. The transmit power of each cell is assumed to be $\bar P_m=20$ dBm and the outage SIR threshold is $\gamma_{\thr}=0$ dB. Other major simulation parameters are summarized in Table~\ref{table:simpars}.

\begin{table}
\small
\centering
\caption{Main simulation parameters.}\label{table:simpars}
\begin{tabular}{p{5cm}|p{1cm}}
\hline
{\bf Simulation parameter}  & {\bf Value}  \\
\hline
Maximum number of episodes $\bar{N}_{\mathrm{epi}}$ & 5000\\
\hline
Initial exploration parameter  $\epsilon_0$ & 0.5 \\
\hline
Exploration decaying rate $\alpha$  & 0.998 \\
\hline
Replay memory capacity $C$  & 100000\\
\hline
Steps for multi-step learning $N_1$ & 30\\
\hline
Update frequency for target network $B$ & 5 \\
\hline
Maximum step per episode $\bar{N}_{\mathrm{step}}$ & 200\\
\hline
Step displacement $\Delta_s$ & 10 m \\
\hline
Reaching-destination reward $R_{\mathrm{des}}$ & 200 \\
\hline
Outbound penalty $P_{\mathrm{ob}}$ & 10000 \\
\hline
Outage penalty weight $\mu$ & 40 \\
\hline
Number of signal measurements per time step $J$ & 1000\\
\hline
\end{tabular}
\vspace{-3ex}
\end{table}

\begin{figure}
\centering
\begin{subfigure}[b]{0.47\textwidth}
\centering
\includegraphics[width=\textwidth]{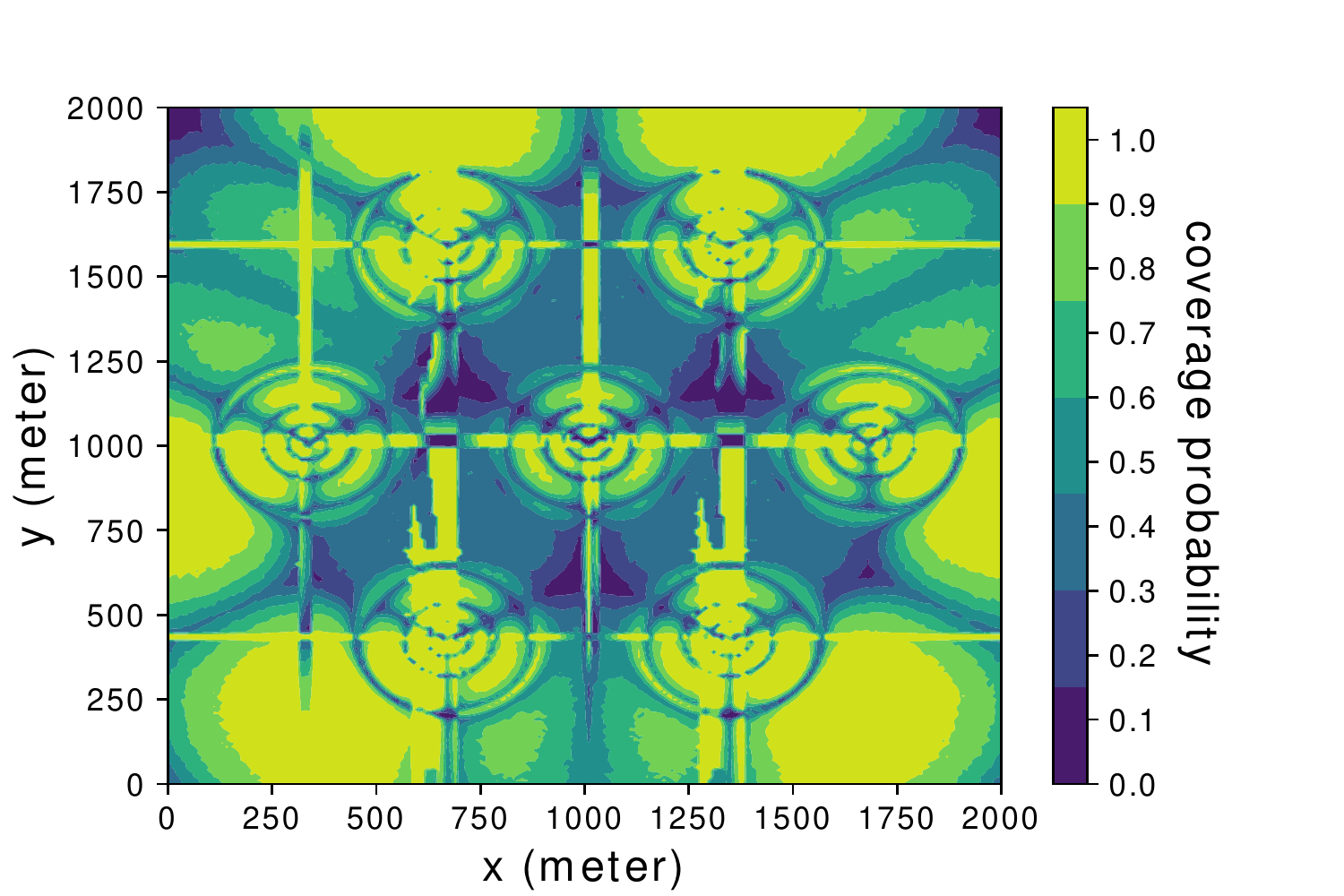}
\caption{True coverage map}
\end{subfigure}
\hfill
\begin{subfigure}[b]{0.47\textwidth}
\centering
\includegraphics[width=\textwidth]{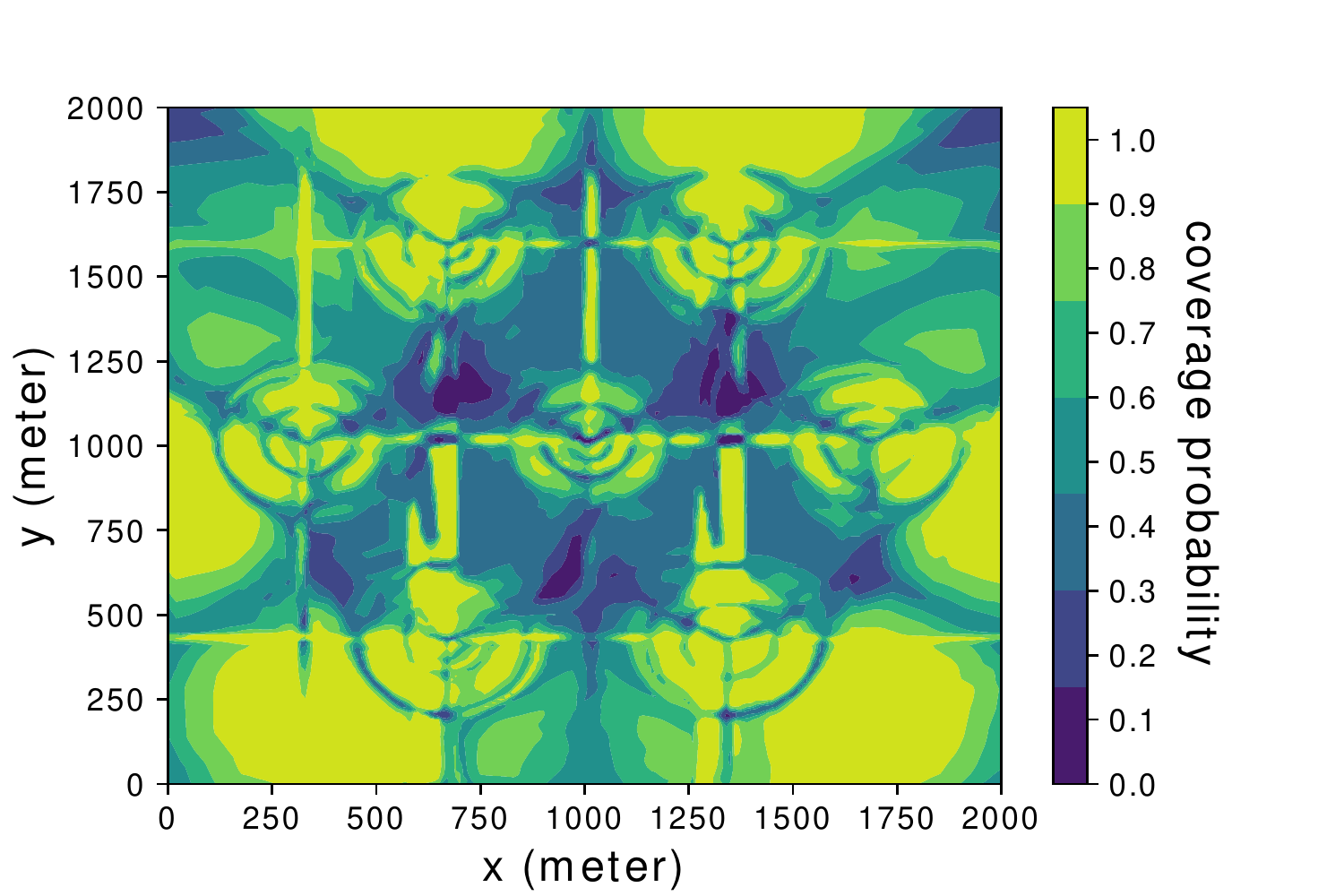}
\caption{Learned coverage map}
\end{subfigure}
\caption{The true  versus  learned coverage maps, where the coverage probability at each location is the complementary probability of the outage probability.}
\label{F:CoverageMap}
\end{figure}

Fig.~\ref{F:CoverageMap}(a) shows the true global coverage map for the considered area, i.e., the coverage (non-outage) probability for each location by the cellular network, where coverage probability is the complementary probability of the outage probability defined in \eqref{eq:Pout}. Note that such a global coverage map is numerically obtained based on the aforementioned 3D building  and channel realizations via computer simulations, while it is not available for our proposed Algorithm~\ref{Algo:PathLearning} or Algorithm~\ref{Algo:SNARM}. It is observed from Fig.~\ref{F:CoverageMap}(a) that the coverage pattern in the sky is rather irregular due to the joint effects of the 3D BS antenna radiation pattern and building blockage. It is observed that around the center of the area, there are several weak coverage regions with coverage probability less than $30\%$. Obviously, effective coverage-aware UAV navigation should direct UAVs to avoid entering such weak coverage regions with the best effort.

\begin{figure}
\centering
\begin{subfigure}[b]{0.47\textwidth}
\centering
\includegraphics[width=\textwidth]{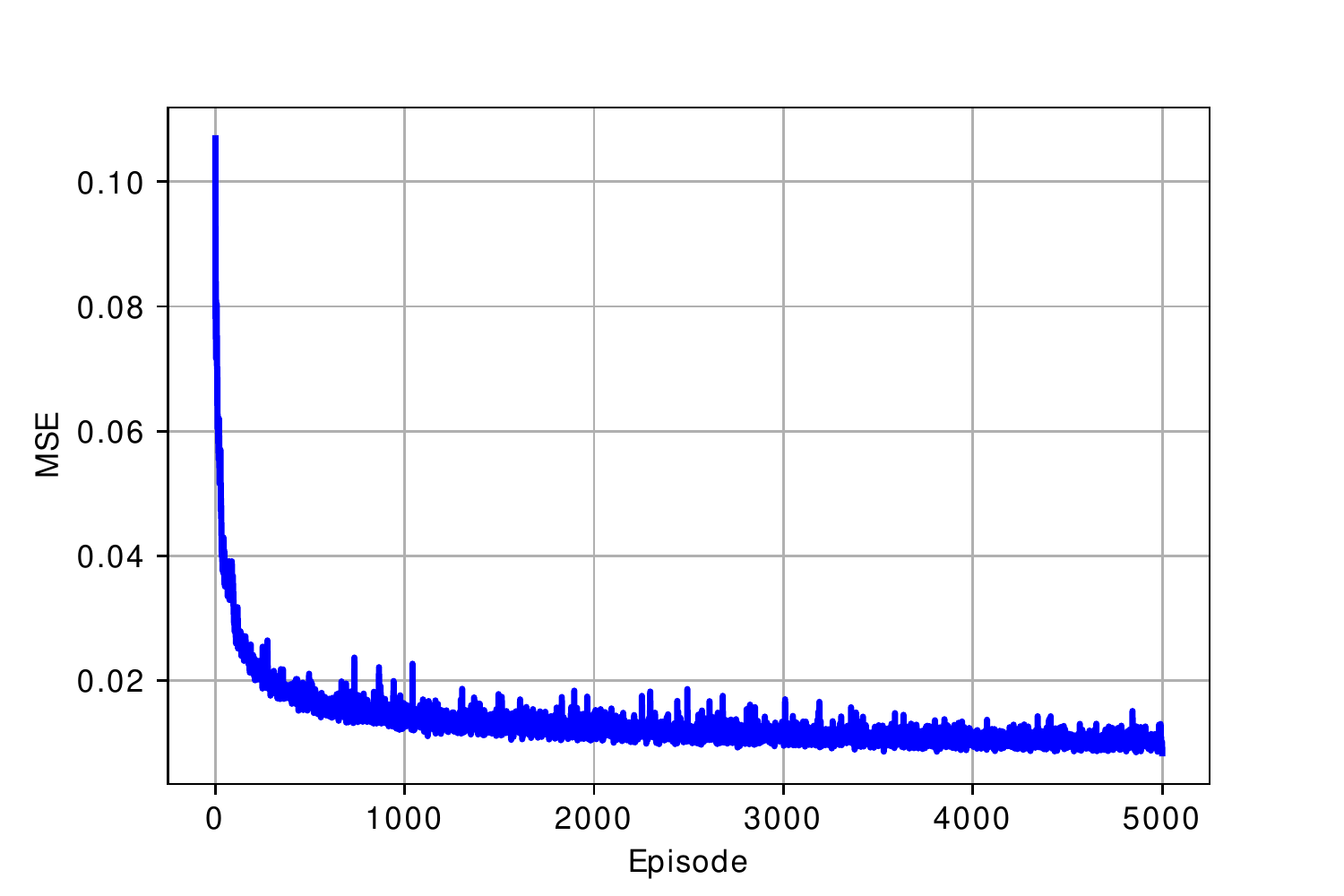}
\caption{MSE}
\end{subfigure}
\hfill
\begin{subfigure}[b]{0.47\textwidth}
\centering
\includegraphics[width=\textwidth]{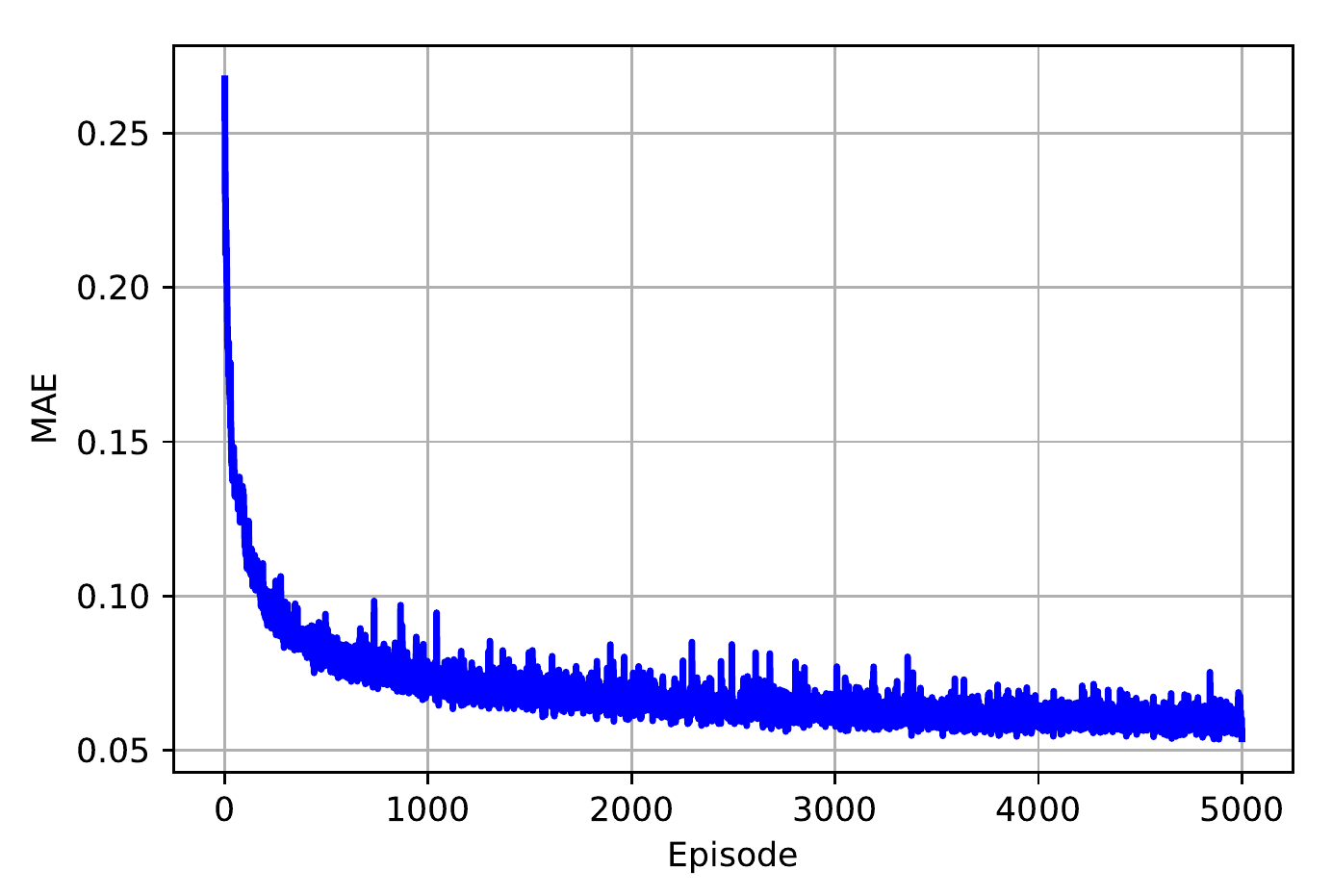}
\caption{MAE}
\end{subfigure}
\caption{The MSE and MAE of radio mapping versus episode number.}
\label{F:MSEandMAE}
\end{figure}

To validate the quality of radio map estimated by the SNARM framework proposed in Algorithm~\ref{Algo:SNARM}, Fig.~\ref{F:CoverageMap}(b) shows the finally learned coverage map by Algorithm~\ref{Algo:SNARM}. By comparing Fig.~\ref{F:CoverageMap}(a) and  Fig.~\ref{F:CoverageMap}(b), it is observed that the learned coverage map is almost identical to the true map, with only slight differences. This thus validates the effectiveness of applying SNARM for the radio map estimation and coverage-aware path learning based on it. This is further corroborated by Fig.\ref{F:MSEandMAE}, which shows  the MSE and mean absolute error (MAE) of the learned radio map versus the episode number. Note that the MSE and MAE are calculated by comparing the predicted outage probabilities using the learned radio map versus their actual values in the true map for a set of randomly selected locations. It is observed that the quality of the learned radio map is rather poor initially, but it improves rather quickly with the episode number as more signal measurements are accumulated. For example, with only 200 episodes, we can already achieve 87.1\% of the MSE reduction that is achieved after 5000 episodes. 

%

\begin{figure}
\centering
\includegraphics[width=0.6\textwidth]{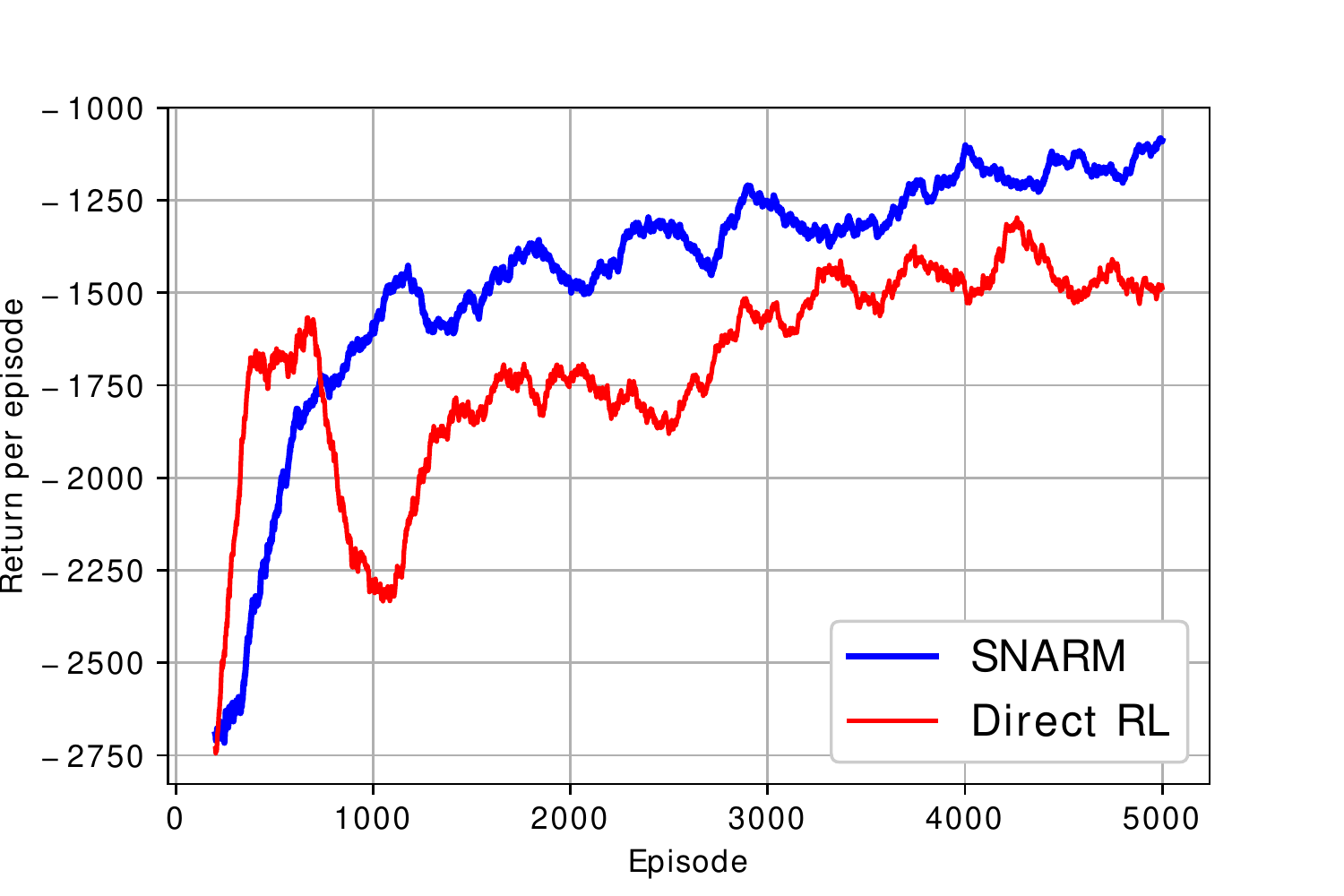}
\caption{Moving average return of direct RL versus SNARM over episode. \vspace{-3ex}}
\label{F:returnComprison}
\end{figure}

As for the performance of the proposed learning algorithms, Fig.~\ref{F:returnComprison} shows the moving average return per episode for the direct RL in Algorithm~\ref{Algo:PathLearning} and SNARM in Algorithm~\ref{Algo:SNARM}, where the moving window has the length of 200 episodes. For the SNARM algorithm, the number of steps $\tilde N$ based on simulated experience per actual step at episode $n$ is set as $\tilde N=\min (\lfloor n/100\rfloor, 10)$, so that $\tilde N$ increases with the episode number and converges to 10. Recall that the return of each episode for RL is the accumulated rewards for all time steps in the episode. It is observed that though experiencing certain fluctuation, as usually the case for RL algorithms, both proposed algorithms demonstrate an overall tendency of increasing average return. Furthermore, recall that the dueling DQNs are initialized to encourage shortest path flying for the early episodes when the UAV has completely no knowledge about the radio environment, which leads to an average return about $-2700$ initially. After training for 5000 episodes, the resulting return has increased to $-1089$ and $-1486$ for SNARM and direct RL, respectively. Another observation from the figure is that the direct RL has faster improvement over the SNARM algorithm at the earlier episodes. This is probably due to the insufficient accuracy of the learned radio map for SNARM, which may generate inefficient simulated experience. However, as the quality of the radio map improves, the SNARM algorithm achieves better performance than direct RL, thanks to its exploitation of the learned radio map for path planning. In practice, such a performance improvement translates to fewer agent-environment interactions, which is highly desirable due to the usually high cost of data acquisition from real experience. For example, to achieve an average return of $-1500$, Fig.~\ref{F:returnComprison} shows that about 3200 actual UAV flight episodes are required for direct RL in Algorithm~\ref{Algo:PathLearning}, but this number is significantly reduced to about 1000 with SNARM in Algorithm~\ref{Algo:SNARM}.

\begin{figure}
\centering
\begin{subfigure}[b]{0.47\textwidth}
\centering
\includegraphics[width=\textwidth]{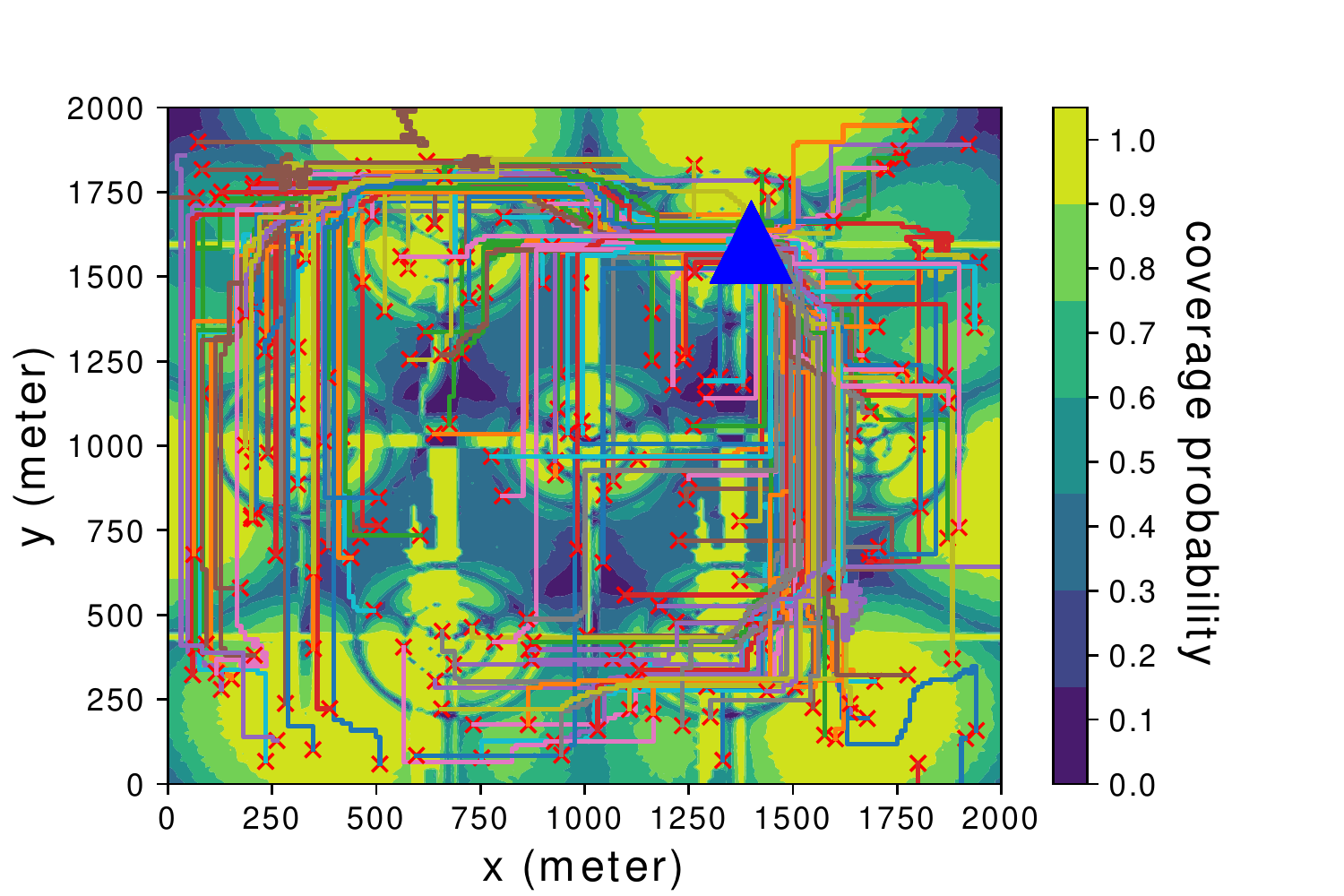}
\caption{Direct RL}
\end{subfigure}
\hfill
\begin{subfigure}[b]{0.47\textwidth}
\centering
\includegraphics[width=\textwidth]{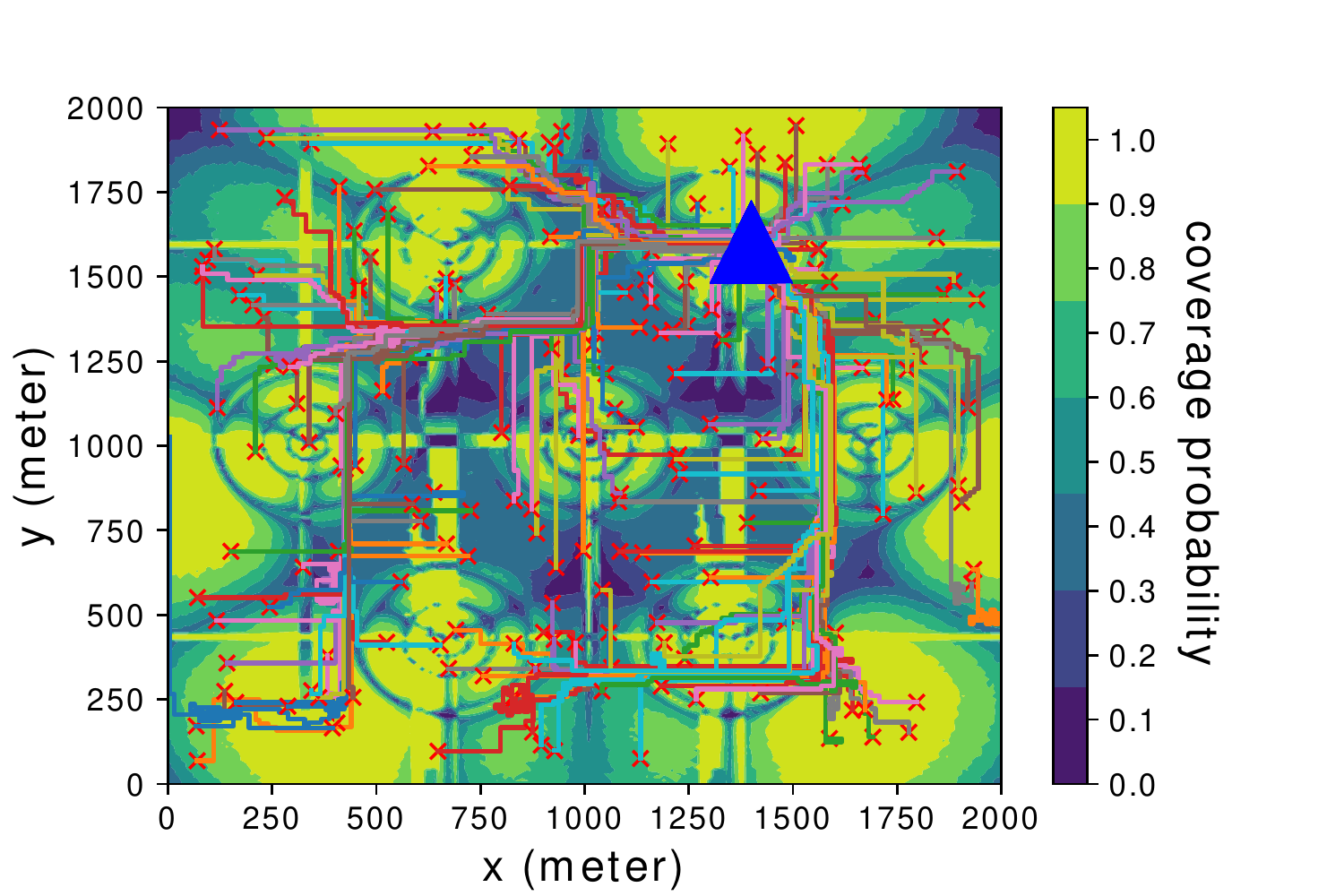}
\caption{SNARM}
\end{subfigure}
\caption{The resulting UAV trajectories with direct RL in Algorithm~\ref{Algo:PathLearning} and SNARM in Algorithm~\ref{Algo:SNARM}. All episodes share a common destination location labelled by the big blue triangle, and their initial locations are randomly generated, labelled by red crosses.}
\label{F:trajectories}
\end{figure}

Last, Fig.~\ref{F:trajectories} shows the resulting UAV paths with the proposed algorithms for the last 200 episodes of learning, which correspond to 200 randomly generated initial locations, as denoted by red crosses in the figure. The common destination location $\mathbf q_F$ is labelled by a big blue triangle.  Also shown in the figure is the true global coverage map. It is observed that both direct RL and SNARM algorithms are able to direct the UAVs to avoid the weak cellular coverage regions with the best effort, as evident from the generally detoured UAV paths as well as the much sparser paths in areas with low coverage probabilities. For example, as can be seen from Fig.~\ref{F:trajectories}(b), the SNARM algorithm is able to discover and follow the narrow ``radio bridge'' located at the x axis around 1000 m and the y axis from about 1000 m to 1700 m. This shows the peculiar effectiveness of SNARM for coverage-aware UAV path learning.

\section{Conclusions}\label{sec:Conclusion}
This paper studies coverage-aware navigation for cellular-connected UAVs. To overcome the practical limitations of the conventional optimization-based path design approaches, we propose DRL-based algorithms, which only require UAV signal measurements as the input. By utilizing the state-of-the-art dueling DDQN with multi-step learning, a direct RL-based algorithm is first proposed, followed by the more advanced SNARM framework to enable radio mapping and reduce the real UAV flights for data acquisition. Numerical results are provided to show the effectiveness of the proposed algorithms for coverage-aware UAV navigation, and the superior performance of SNARM over direct RL based navigation.

\bibliographystyle{IEEEtran}
\bibliography{IEEEabrv,IEEEfull}

\end{document}